\documentclass[prx,superscriptaddress,twocolumn,10pt,aps,floatfix]{revtex4-1}

\input{glyphtounicode}
\usepackage[T1]{fontenc}
\usepackage[utf8]{inputenc}
\usepackage[english]{babel}
\usepackage{amsmath} 
\usepackage{amsfonts}
\usepackage{graphicx} 
\usepackage[pdftex,bookmarks=true,bookmarksopen,bookmarksnumbered,
                colorlinks,
                linkcolor=blue,
                citecolor=blue]{hyperref}
\graphicspath{ {./} }
\usepackage{natbib}
\usepackage{array}
\usepackage[table]{xcolor}
\usepackage{xcolor}
\usepackage{braket}
\usepackage{caption}

\usepackage{booktabs}
\usepackage{multirow}

\newcommand{\aref}[1]{\hyperref[#1]{Appendix~\ref*{#1}}}

\begin{document}

\captionsetup[table]{name={TABLE},labelsep=period,justification=raggedright,font=small}
\captionsetup[figure]{name={FIG.},labelsep=period,justification=raggedright,font=small}
\renewcommand{\equationautorefname}{Eq.}
\renewcommand{\figureautorefname}{Fig.}
\renewcommand*{\sectionautorefname}{Sec.}

\title{Implementation of the SMART protocol for global qubit control in silicon}

\author{Ingvild Hansen}\affiliation{School of Electrical Engineering and Telecommunications, The University of New South Wales, Sydney, NSW 2052, Australia}
\author{Amanda E. Seedhouse}\affiliation{School of Electrical Engineering and Telecommunications, The University of New South Wales, Sydney, NSW 2052, Australia}
\author{Kok Wai Chan}\affiliation{School of Electrical Engineering and Telecommunications, The University of New South Wales, Sydney, NSW 2052, Australia}
\author{Fay Hudson}\affiliation{School of Electrical Engineering and Telecommunications, The University of New South Wales, Sydney, NSW 2052, Australia}
\author{Kohei M. Itoh}\affiliation{School of Fundamental Science and Technology, Keio University, Yokohama, Japan}
\author{Arne Laucht}\affiliation{School of Electrical Engineering and Telecommunications, The University of New South Wales, Sydney, NSW 2052, Australia}
\author{Andre Saraiva}\affiliation{School of Electrical Engineering and Telecommunications, The University of New South Wales, Sydney, NSW 2052, Australia}
\author{Chih Hwan Yang}\affiliation{School of Electrical Engineering and Telecommunications, The University of New South Wales, Sydney, NSW 2052, Australia}
\author{Andrew S. Dzurak}\affiliation{School of Electrical Engineering and Telecommunications, The University of New South Wales, Sydney, NSW 2052, Australia}
\date{\today}

\begin{abstract}
Quantum computing based on spins in the solid state allows for densely-packed arrays of quantum bits. While high-fidelity operation of single qubits has been demonstrated with individual control pulses, the operation of large-scale quantum processors requires a shift in paradigm towards global control solutions.
Here we report the experimental implementation of a new type of qubit protocol -- the SMART (Sinusoidally Modulated, Always Rotating and Tailored) protocol. As with a dressed qubit, we resonantly drive a two-level system with a continuous microwave field, but here we add a tailored modulation to the dressing field to achieve increased robustness to detuning noise and microwave amplitude fluctuations. We implement this new protocol to control a single spin confined in a silicon quantum dot and confirm the optimal modulation conditions predicted from theory. Universal control of a single qubit is demonstrated using modulated Stark shift control via the local gate electrodes. We measure an extended coherence time of $2$\,ms and an average Clifford gate fidelity $>99$\,$\%$ despite the relatively long qubit gate times ($>15$\,\textmu s, $20$ times longer than a conventional square pulse gate), constituting a significant improvement over a conventional spin qubit and a dressed qubit. This work shows that future scalable spin qubit arrays could be operated using global microwave control and local gate addressability, while maintaining robustness to relevant experimental inhomogeneities.

\end{abstract}

\pacs{}

\maketitle

\section{Introduction}

The successful implementation of full scale quantum computing is expected to drive technological progress in a multitude of areas, however, the realisation of a fault-tolerant quantum computer requires qubits that are resilient to errors. Most qubit implementations will require Quantum Error Correction (QEC), which sets strict requirements on qubit fidelities \cite{bennett1996mixedstate,devitt2013quantum}. At most an error rate of $1\,\%$ is tolerable for a two-dimensional qubit array using the surface code \cite{knill2005quantum,fowler2012surface}. Depending on the error rate, the number of physical qubits is expected to exceed millions, underpinning the need for the qubits to be not only highly robust against errors but also readily scalable. Silicon quantum dots, in particular those based on silicon metal–oxide–semiconductor (SiMOS) devices, are promising candidates for qubits due to their long coherence times and their compatibility with current semiconductor manufacturing capabilities \cite{Morton2011Embracing,veldhorst2014addressable,veldhorst2015twoqubit, veldhorst2017silicon,Zhang2018Semiconductor}.  However, even with isotopically enriched silicon substrates, residual nuclear spins \cite{hensen2020silicon,zhao2019singlespin} and spin-orbit-coupling due to interface disorder \cite{veldhorst2015spinorbit,ruskov2018electron}  reduce both the coherence time and the homogeneity of the spin qubit properties. It is therefore important to have a control scheme that is resilient to qubit variability.

Different global qubit control strategies, amenable to large-scale operation and manufacturing, have been discussed in recent works \cite{vahapoglu2021singleelectron,vallabhapurapu2021fast,seedhouse2021quantum,vahapoglu2021coherent}. A control strategy proposed by Kane in 1998 \cite{kane1998siliconbased} involves a global microwave field that is by default off-resonance with the qubits, and local controls are applied to tune the individual qubits into resonance to perform gate operations \cite{laucht2015electrically}. These qubits, however, can suffer from reduced coherence due to environmental noise. Another option is to collectively drive all qubits on-resonance by default, to operate them as dressed qubits \cite{seedhouse2021quantum, golter2014protecting,wu2019adiabatic,mikelsons2015universal}.  The global dressing field provides continuous decoupling of the two-level systems from the noisy environment, enabling longer coherence times \cite{laucht2017dressed, Miao2020Universal}.  For scalability this scheme is, however, limited by the spread in resonance frequencies of the qubits, making them susceptible to crosstalk \cite{jones2018logical}. To make dressed qubits more robust to frequency deviations (typically around $200$\,kHz for quantum dot spin qubits \cite{hensen2020silicon}) and amplitude fluctuations of the driving field, a new type of qubit dressing, named the SMART protocol, has been proposed \cite{hansen2021smart}, which employs a custom-designed modulation of the global microwave field.

\begin{figure*}[htb]
    \centering
    \includegraphics[width = 1\textwidth,angle=0]{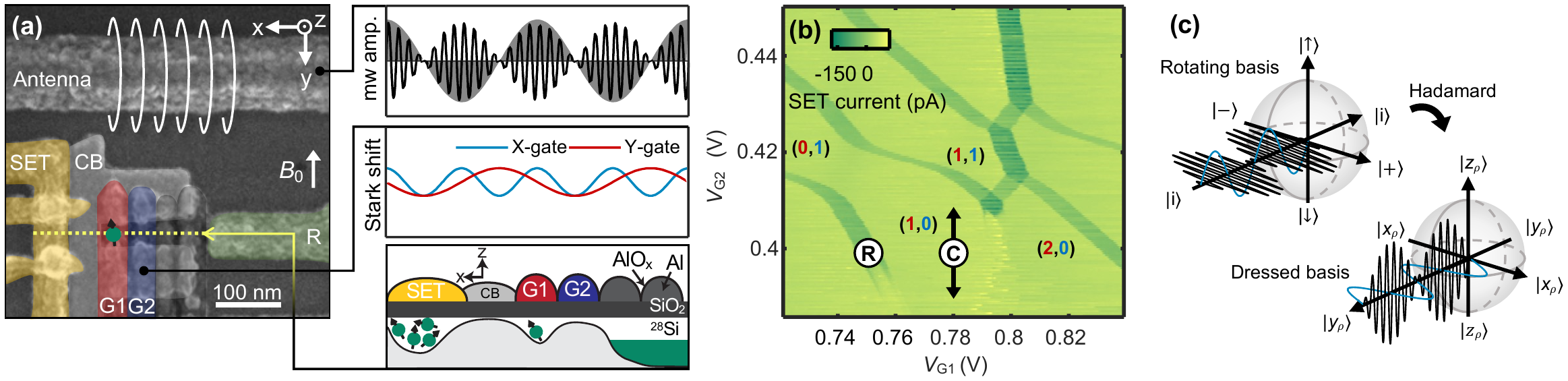}
    \caption{(a) False-colored scanning electron micrograph of an identical device to that used in this experiment. A quantum dot is formed in a silicon MOS structure underneath the G1 gate, laterally confined by the confinement barrier (CB) with electrons provided by the reservoir (R). A single electron transistor (SET) under gate SET is used for charge sensing. The arrow indicates the direction of an in-plane external magnetic field \cite{yoneda2021coherent}. An IQ-modulated microwave source provides the modulated, oscillating magnetic driving field via the antenna (top right inset), and $V_{\rm{G2}}$ is pulsed for qubit control (centre right inset). Schematic cross-section of the device at the position of the dashed line in panel (a) (bottom right inset). 
    (b) Charge stability map showing the control point (C) and the readout point (R). The arrows show the movement in voltage space during pulsing of $V_{\rm{G2}}$ for qubit control. 
    (c) Transformation from the rotating basis to the dressed basis with the microwave modulation (black) and the Stark shift modulation (blue). 
    }
    \label{fig:device}
\end{figure*}

Here we use a SiMOS quantum dot device to confirm the increased robustness of qubits operated using the SMART protocol to time dependent detuning and microwave amplitude fluctuations, as suggested in Ref. \onlinecite{hansen2021smart}. The experiments are performed on a device identical to the one shown in \autoref{fig:device}(a) and investigated in earlier work \cite{yoneda2021coherent,yang2019silicon,chan2018assessment}. A single quantum dot is formed under gate G1, while gate G2 is pulsed to control the qubit through Stark shift of the gyromagnetic ratio \cite{veldhorst2014addressable}. The experiment is conducted in a dilution refrigerator with an electron temperature of $100-150$ mK. We confirm the optimal modulation conditions from theoretical predictions, and demonstrate coherent and universal control as well as one qubit randomised benchmarking.

\section{The SMART qubit protocol}
In this work we are investigating a qubit continuously driven by a resonant a.c. magnetic field that is amplitude-modulated by a sinusoid. For this kind of driven qubit we work in the dressed basis, in which the quantisation axis is defined perpendicular to the $|{\uparrow}\rangle$/$|{\downarrow}\rangle$ axis and the basis states are the superposition states $|+\rangle=\frac{1}{\sqrt{2}}(|{\downarrow}\rangle+|{\uparrow}\rangle)$ and $|-\rangle=\frac{1}{\sqrt{2}}(|{\downarrow}\rangle-|{\uparrow}\rangle)$  \cite{laucht2017dressed}. This implies the following axes transformations from the rotating basis: $|{\uparrow}\rangle\rightarrow|x_{\rho}\rangle$, $|{\downarrow}\rangle\rightarrow|\bar{x}_{\rho}\rangle$, $|+\rangle\rightarrow|z_{\rho}\rangle$, $|-\rangle\rightarrow|\bar{z}_{\rho}\rangle$, $|i\rangle\rightarrow|\bar{y}_{\rho}\rangle$ and $|\bar{i}\rangle\rightarrow|y_{\rho}\rangle$ [see \autoref{fig:device}(c)]. Notice how the sign of dressed basis $y_{\rho}$ is flipped compared to the rotating basis. The same basis transformation is used in Ref. \onlinecite{seedhouse2021quantum,hansen2021smart}. The SMART protocol qubit system is defined by the Hamiltonian in the dressed basis $\{\ket{z_\rho}, \ket{\bar{z}_\rho}\}$

\begin{equation}
        H^{\rm{cos}}_\rho = \frac{h}{2}\bigg(\Omega_{\rm{R}}\sqrt{2}\cos(2\pi{f}_{\text{mod}}t)\sigma_z+\Delta\nu(t)\sigma_x\bigg),
        \label{eq:h}
\end{equation}
where $h$ is the Planck constant, $\Omega_{\rm{R}}$ the Rabi frequency, $f_{\text{mod}}$ the modulation frequency and $\Delta\nu(t)$ the frequency detuning (see \aref{app:Hd} for derivation of $H^{\rm{cos}}_\rho$ from the laboratory frame). One period of the sinusoidal modulation will from now on be denoted $T_{\rm{mod}}$. The $\sqrt{2}$ factor  accounts for the root mean square power difference between a square pulse and a sinusoid. Similarly to \autoref{eq:h} we can modulate the global field with a sine instead of a cosine. We will refer to these two variants as SMART (cos) and SMART (sin).

Measurement of qubits in the SMART qubit protocol is based on initializing a $|{\downarrow}\rangle$ state and reading out the $|{\uparrow}\rangle$ probability using a single electron transistor (SET). To transform into the dressed basis we apply a resonant $\pi/2$ pulse about $-y$ rotating the $|{\downarrow}\rangle$ state into the $|+\rangle$ state \cite{laucht2017dressed}. Spin-to-charge conversion is used for spin readout \cite{elzerman2004singleshot} at the first electron transition [\autoref{fig:device}(b)]. 

All controls applied to the device for the SMART qubit protocol are a.c., hence one can use high/band-pass filters to reduce d.c.- and high frequency noise. For SMART (cos) operation we add a pulse sequence at the beginning and end of the microwave sequence to avoid sharp jumps in amplitude (see \aref{app:amp}).

\begin{figure*}
    \centering
    \includegraphics[width = 1\textwidth, angle = 0]{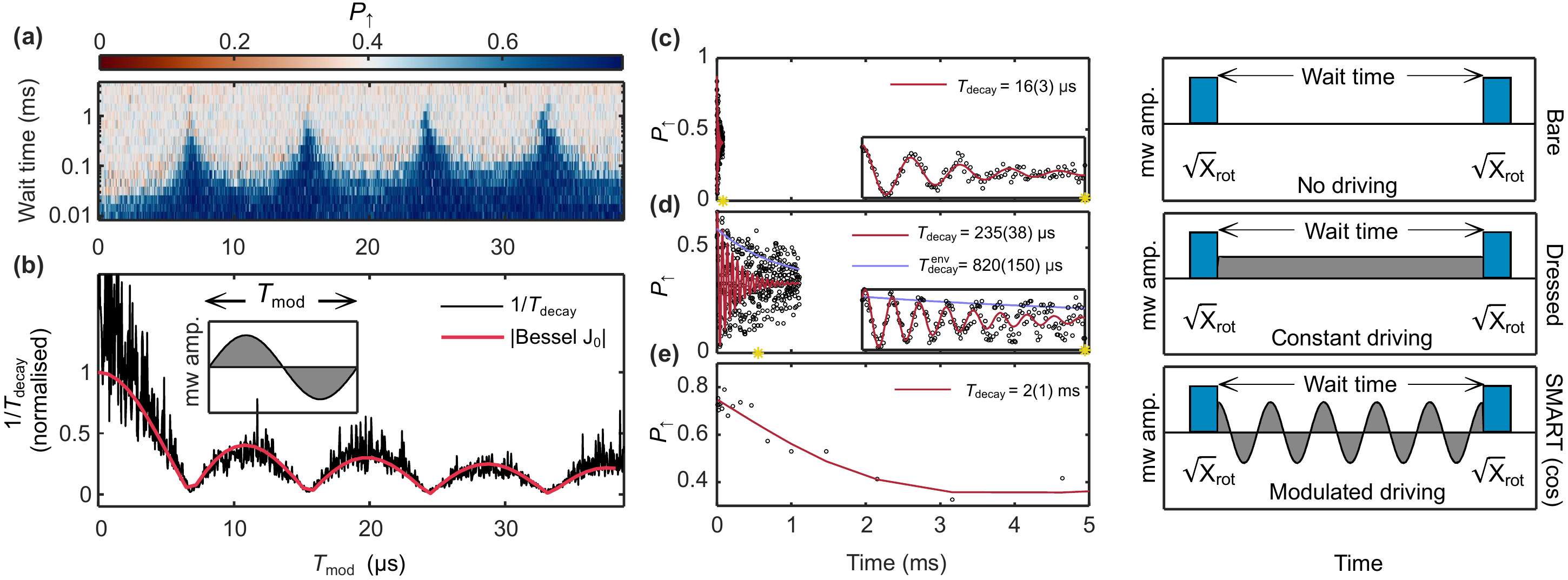}
    \caption{(a) A SMART (cos) protocol Ramsey experiment where the period of the global modulation is varied showing that there exist optimal modulation conditions. The spin is initialised to $|\bar{y}_{\rho}\rangle$ and projected onto the $|{\uparrow}\rangle$ axis with a microwave square pulse after wait time $t_{\rm{wait}}$. At wait time $0.6$\,ms  a fluctuation in MW output power distorts the data. The data is taken with $300$ spin shots repeated three times. (b) The extracted decay rates (black line) are fitted to the absolute value of the Bessel function of zeroth order (red line). (c-e) Comparison between bare ($100$ shots, six repeats), dressed ($300$ shots, two repeats) and SMART protocol (at $T_{\rm{mod}}=24.28\,$\textmu s) qubit decay times (left column), together with the respective microwave pulse sequences (right column). Here $\sqrt{\rm{X}}_{\rm{rot}}$ is in the rotating basis initialising the spin along $|{i}\rangle=|{\bar{y}_{\rho}}\rangle$. 
    }
    \label{fig:ramsey}
\end{figure*}

\section{Coherence times}

The coherence time of a driven spin qubit cannot be directly compared to that of a bare spin qubit as they are not sensitive to the same type of noise.  Driven spin qubits are dominated by noise at the frequency of the driving field \cite{yan2013rotatingframe,hansen2021smart}. The quantisation axis ($T_1$ axis) for the dressed qubit is along the driving field axis $z_{\rho}$. For the SMART (cos) protocol on the other hand, the qubit quantisation axis is along $x_{\rho}$ and along the $w$-axis for the SMART (sin) protocol. Due to this we define a new parameter $T_{\rm{decay}}$ as the measured coherence time of the spin rotation resulting from the driving field, when initialised to $|\bar{y}_{\rho}\rangle$ (driving field along $z_{\rho}$). We note that more extensive noise spectroscopy/tomography should be completed to get a better picture of the noise mechanics in driven qubit systems. This is left for future work.

After initialising the qubit to $|\bar{y}_{\rho}\rangle$ via a $\pi/2$ rotation about $z_{\rho}$ (denoted $\sqrt{\text{X}}_{\text{rot}}$ as it is a rotation about $x$ in the rotating basis and $z_{\rho}$ in the dressed basis) the modulated driving about the same axis is turned on for a certain wait time, $t_{\text{wait}}$, followed by a final $\pi/2$ rotation about $z_{\rho}$ [$\sqrt{\rm{X}}_{\rm{rot}}-t_{\text{wait}}-\sqrt{\rm{X}}_{\rm{rot}}$, see \autoref{fig:ramsey}(e) right panel]. The spin is then measured at increasing multiples of $T_{\text{mod}}$, where the modulated driving itself ideally equates to an identity operation after every period. By keeping $\Omega_{\rm{R}}$ fixed and varying the period $T_{\rm{mod}}$, the 2D map in \autoref{fig:ramsey}(a) is acquired. Similarly, $T_{\rm{mod}}$ can be kept fixed while varying $\Omega_{\rm{R}}$. The data is fitted according to an exponential decay $\propto e^{(-t/T_{\rm{decay}})}$, resulting in the decay rate plotted in \autoref{fig:ramsey}(b). The decay rate as a function of $T_{\text{mod}}$ closely resembles the absolute value of the zeroth order Bessel function plotted for comparison. The maximum $T_{\rm{decay}}$ measured here is $2(1)$\,ms at $T_{\rm{mod}}=24.28$\,\textmu s. For comparison the bare spin qubit ($T_2^*$) and the dressed qubit coherence times are measured to be $16(3)$\,\textmu s and $235(38)$\,\textmu s in the same device [\autoref{fig:ramsey}(c-d)]. The dressed data is very sensitive to changes in precession frequency, therefore a fit of the envelope is shown as well. For the remaining part of this paper we will focus on the first peak from left in \autoref{fig:ramsey}(a) %corresponding to $j_1$ in \autoref{eq:opt}
where $T_{\text{mod}}=6.76$\,\textmu{s} and $T_{\text{decay}}= 1.6(6)$\,ms.

The improvement in coherence time with distinct periods of the microwave field modulation, as apparent by the four peaks in \autoref{fig:ramsey}(a), agrees with the theoretical prediction originating from the geometric formalism in \cite{zeng2019geometric,yang2019silicon}. Certain modulation frequencies are more desirable because they cancel out both first and second order noise, making the SMART qubit protocol more robust to detuning and microwave amplitude noise. From  \autoref{fig:ramsey}(a-b) it can be seen how the optimal modulation period \cite{hansen2021smart} follows

\begin{equation}
    {T}^{\text{opt}}_{\text{mod}}= j_i/\Omega_{\rm{R}}\sqrt{2},
    \label{eq:opt}
\end{equation}
where $j_i$ is solution $i$ to the zeroth order Bessel function. In this work we focus on the solution with $j_1 = 2.404826$ corresponding to the first peak in \autoref{fig:ramsey}(a).

The width of the four peaks in \autoref{fig:ramsey}(a) demonstrates the robustness to microwave strength spatial variation. One can get significant amplitude variation if the antenna is not broadband, for example if a microwave resonator is used to provide the driving field to the spins. Another limitation is set by the global field modulation frequency. A $6$\,GHz resonator with a Q-factor of $10^4$ only has a $600$\,kHz bandwidth. The resonator bandwidth must be in the range of the modulation frequency (MHz) in order to successfully create qubits with the SMART protocol. This can be relaxed by choosing a larger $T_{\rm{mod}}$ [peaks to the right in \autoref{fig:ramsey}(a)] at the expense of longer gate times. Another alternative is to use multimode cavities.

\section{More advanced driving fields}

\begin{figure*}
    \centering
    \includegraphics[width  = 1\textwidth]{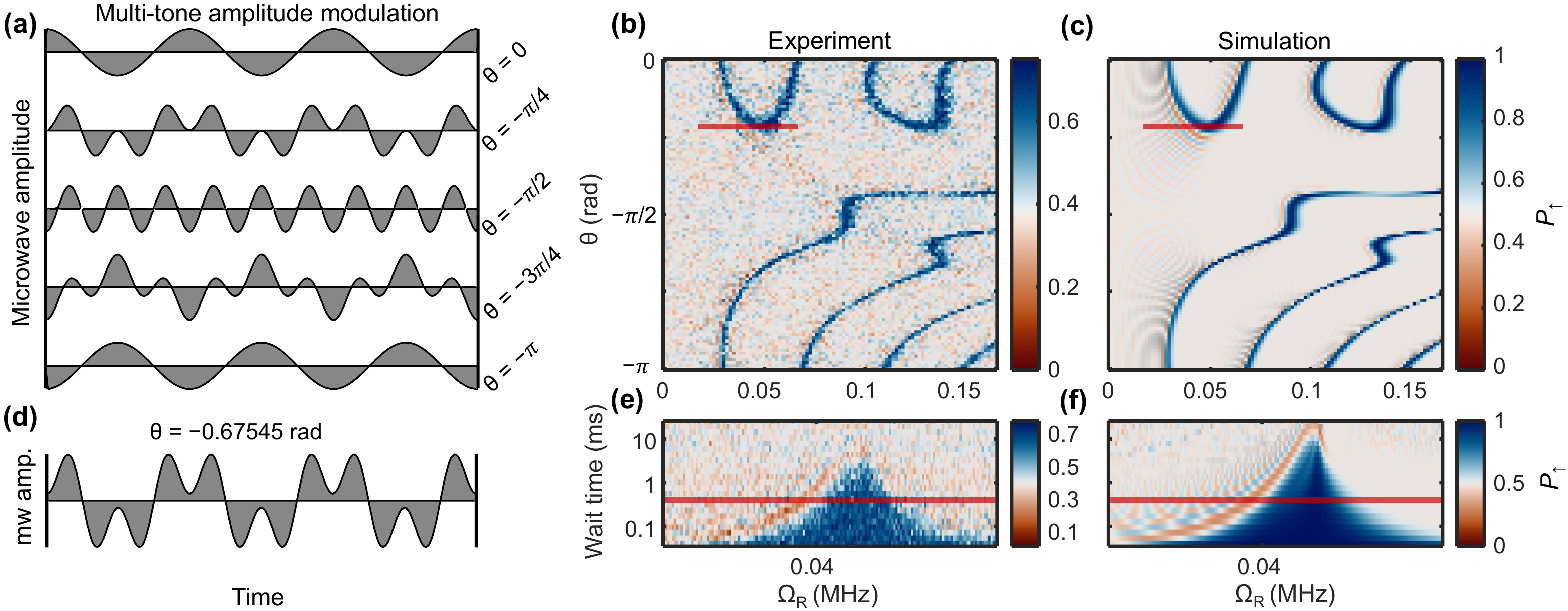}
    \caption{Multi-tone global driving fields. 
    (a) By combining the first and third harmonic of a sinusoid according to \autoref{eq:cd}, driving fields with different properties can be generated. 
    (b)  Experimental Ramsey data ($100$ shots, three repeats) where the wait time is fixed at $400$\,\textmu s and $T_{\text{mod}}=40$\,\textmu s. The $|{\uparrow}\rangle$ probability is shown for different global amplitudes $\Omega_{\rm R}$ and ratios between the first and third harmonic represented by $\theta$. The simulation is shown in (c). (d) The driving field for $\theta=0.67545$ radians. (e) The $|{\uparrow}\rangle$ probability for different wait times and global amplitudes at $\theta=-0.67545$ radians indicated with a red line in panel (b). The red line indicates the $400$\,\textmu s wait time for which the data in (b) was taken. In (f) the corresponding simulated data is plotted.
    }
    \label{fig:cd}
\end{figure*}

We employ a simple single-tone sinusoidal driving field for most measurements in this work, however, more complex modulation shapes can potentially contribute to higher robustness. In \autoref{fig:cd}(b,c) we investigate, experimentally and by simulations, a multi-tone driving field using the first and third harmonic of a cosine modulated microwave field according to

\begin{equation}
D(t)= \Omega_{\rm R}\sqrt{2}[\cos(\theta)\cos(2\pi{f}_{\text{mod}}t)+\sin(\theta)\cos(6\pi{f}_{\text{mod}}t)], 
\label{eq:cd}
\end{equation}
where $\Omega_{\rm R}$ is the driving field amplitude and $\theta$ determines the ratio of the two harmonics, while keeping the total power fixed. We name this SMART (3rd) and examples of different modulation shapes are shown in \autoref{fig:cd}(a). 

The experimental data in \autoref{fig:cd}(b) is acquired in a similar fashion to the Ramsey experiment in  \autoref{fig:ramsey}, but at a fixed wait time of $400$\,\textmu s and $T_{\rm{mod}}=40$\,\textmu s, for a range of driving field amplitudes $\Omega_{\rm R}$ and ratios of the two tones as defined by $\theta$ in \autoref{eq:cd}. The four peaks observed in \autoref{fig:ramsey} correspond to the case of $\theta=0$ for which the relative amplitude of the third harmonic is zero. The high spin-up probability regions ($T_{\rm{decay}}$ $>400\,$\textmu s) in \autoref{fig:cd}(b) indicate the more ideal modulation parameters for the multi-tone drive for which the qubit is less sensitive to detuning and amplitude fluctuations. These regions closely resemble the simulated data in \autoref{fig:cd}(c). 

One of these regions of high robustness is at $\theta = -0.67545$\,rad as indicated by the red lines in \autoref{fig:cd}(b-c), and we show the corresponding modulation shape in \autoref{fig:cd}(d). We further investigate this special case by recording Ramsey data for a range of amplitudes. We plot the experimental data in \autoref{fig:cd}(e) and the corresponding simulations in \autoref{fig:cd}(f). The maximum $T_{\rm{decay}}$ in \autoref{fig:cd}(e) is $2.15$\,ms. The width of the peak represents high resilience to amplitude fluctuations at the order of $10$\% of $\Omega_{\rm R}$. This is an improvement compared to the peaks in \autoref{fig:ramsey}(a).

The use of even more complex or optimally-shaped, arbitrary driving fields might result in even better performance, however we leave the development of the appropriate control terms for future work.

\begin{figure}
    \centering
    \includegraphics[width = 1\columnwidth]{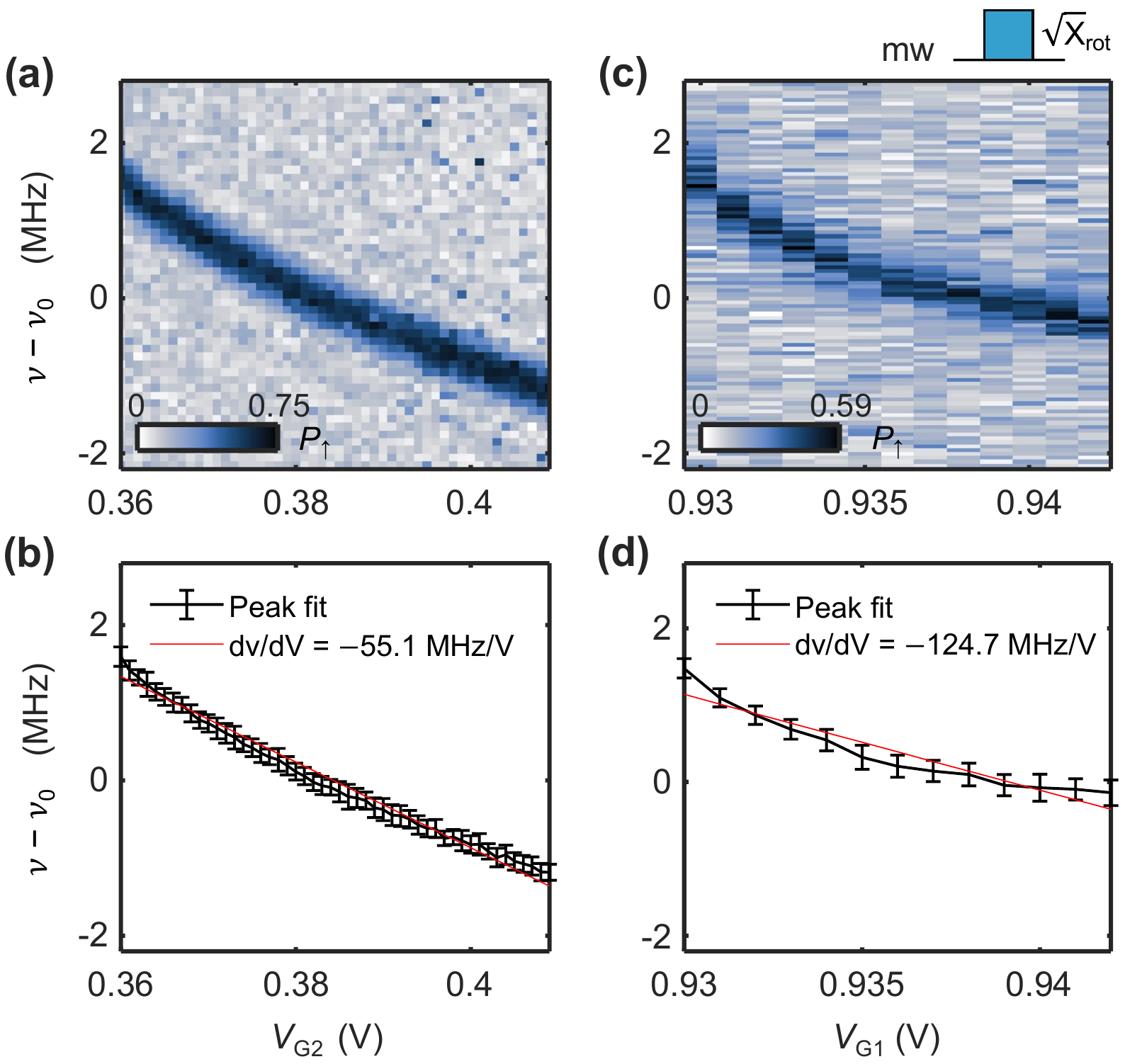}
    \caption{(a) Stark shift of the electron spin resonance as a function of G2 voltage obtained by a square $\pi$-pulse. 
    (b) Fit of the extracted frequency, resulting in a Stark shift magnitude of $-55$\,MHz/V. In (c) and (d) the same is shown for G1 where a Stark shift magnitude of $-125$\,MHz/V is measured.
    }
    \label{fig:stark}
\end{figure}

\begin{figure*}
    \centering
    \includegraphics[width=1\textwidth]{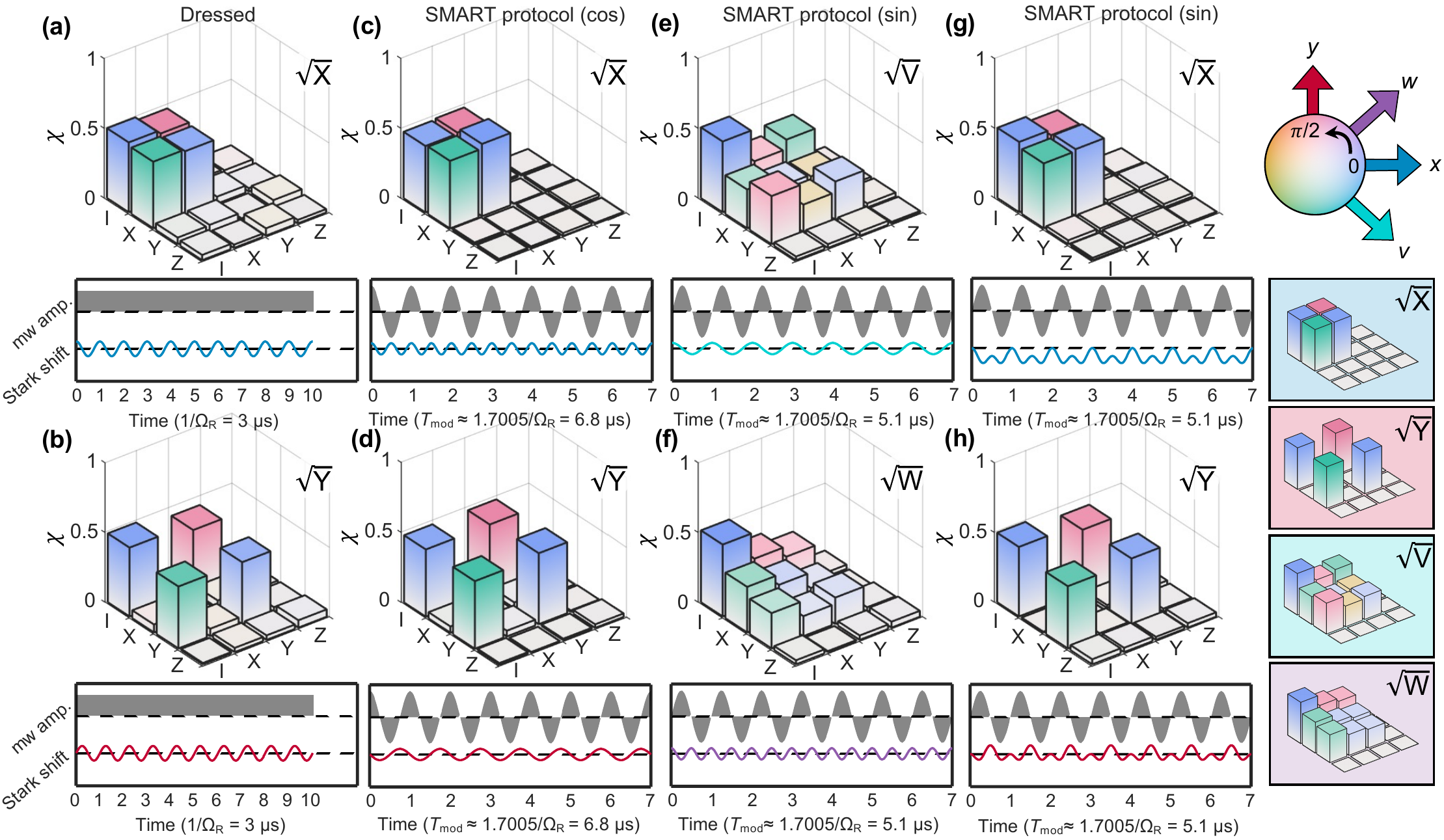}
    \caption{Process tomography of the dressed and SMART qubit protocol showing two-axis control. In (a,b) and (c,d) dressed and SMART (cos) protocol $\sqrt{\rm{X}}$ and $\sqrt{\rm{Y}}$ gates are shown, respectively. In (e-h) $\sqrt{\rm{V}}$,$\sqrt{\rm{W}}$,$\sqrt{\rm{X}}$ and $\sqrt{\rm{Y}}$ gates for SMART (sin) protocol are shown. The height of the bars and the colour code represent the absolute value of the superoperator matrix elements and complex phase information, respectively. On the far right the rotation axes and the ideal superoperators are shown. The data is taken with $120$ spin shots and $\leq 30$ repeats. 
    }
    \label{fig:tomo}
\end{figure*}

\section{SMART protocol qubit control and process tomography}
In order to perform controlled rotations using the SMART qubit protocol, we employ modulated Stark shift control of the spin via the top gate, according to the theory developed in \cite{hansen2021smart,laucht2017dressed}. A typical Stark shift region for the device is shown in \autoref{fig:stark}(a,b) where we measure a Stark shift magnitude of $-55$\,MHz/V for gate G2. For gate G1, we find the Stark shift to have a larger magnitude, but to be less linear. Therefore, we use G2 for gate control throughout this paper.

\begin{figure*}
    \centering
    \includegraphics[width = \textwidth,angle = 0]{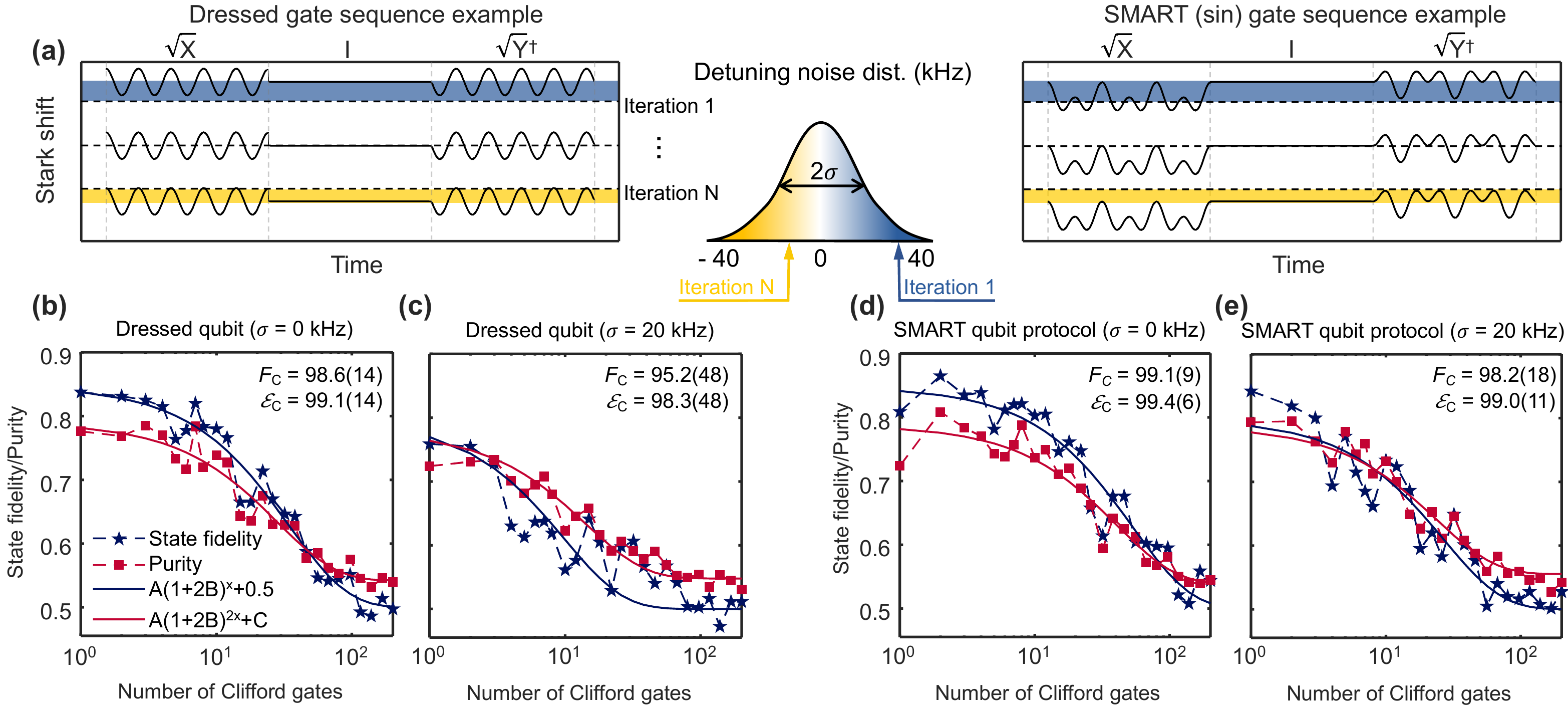}
    \caption{Single qubit randomised benchmarking with added detuning noise for robustness test. In (a) the noise implementation and the respective pulse sequences are shown, where iteration $1$ and N represent positive and negative noise offset, respectively. Randomised benchmarking data is shown for the dressed qubit (b) with no added noise and (c) with $\sigma=20$\,kHz white, quasi-static Gaussian noise added on G2. The same is shown in (d,e) for the SMART (sin) protocol. The state fidelity data is fitted to $A(1+2B)^x+1/2$  and the noise coherence to $A(1+2B)^{2x}+C$. The data is taken with $120$ spin shots and $20$ different Clifford sequences for each sequence length. The gate lengths are here $5/\Omega_{\rm R}$ and $3\times T_{\rm mod}$ for the dressed and SMART case, respectively, where $\Omega_{\rm R}=0.33$\,MHz and $T_{\rm mod}=5.15$\,\textmu{s}.
    }
    \label{fig:rbm}
\end{figure*}

To confirm the rotation axes predicted in \cite{hansen2021smart}, we perform process tomography \cite{nielsen2010quantum}. In order to completely reconstruct the 2 x 2 density matrix, 6 tomography projections are acquired \cite{yang2019silicon} (see \aref{app:atomo} for more details). We demonstrate the two variants of the SMART qubit protocol; SMART (cos) with cosine modulation and SMART (sin) with sine modulation, and compare them to the dressed qubit. In \autoref{fig:tomo}, we show the results for the gates $\sqrt{\rm{X}}, \sqrt{\rm{Y}}, \sqrt{\rm{V}}$ and $\sqrt{\rm{W}}$  for (a,b) the dressed qubit (using FM resonance control \cite{laucht2017dressed}), (c,d) SMART (cos) and (e-h) SMART (sin) protocol. The $\sqrt{\rm{V}}$ and $\sqrt{\rm{W}}$ gates constitute rotations about an alternative, diagonal set of rotation axes (see top right insert), that can be used for the SMART qubit protocol \cite{hansen2021smart}. The individual panels contain the measured superoperator matrices as well as details on the pulse sequences and modulation shapes. For comparison, we plot the ideal superoperators to the far right. All measured superoperator matrices are in agreement with the ideal matrices with good fidelity.

In order for the comparison between the dressed and the SMART qubit protocol to be fair, the same global field root mean square power is used. For controlled rotations, Stark shift amplitudes are chosen such that the gate durations of the two qubits are approximately the same. Here we use gate times for the SMART qubit protocol of $7\times T_{\rm{mod}}$ and for the dressed qubit $10/\Omega_{\rm{R}}$, as shown in \autoref{fig:tomo}. The rotating wave approximation (RWA) must be taken into account here, as discussed in Refs.
\onlinecite{laucht2016breaking,hansen2021smart}.

\section{Randomised benchmarking}
In order to assess the performance of the SMART qubit protocol we carry out randomised benchmarking. Here we determine the average Clifford gate fidelity $F_{\rm{C}}$ and the noise coherence $\xi_{\rm{C}}$ \cite{feng2016estimating} according to the state purity. The 24 Clifford gates are generated using the dressed basis gate set $\{\rm{X},\rm{Y},\pm\sqrt{\rm{X}},\pm\sqrt{\rm{Y}}\}$. The results for the dressed and SMART qubit protocol are presented in \autoref{fig:rbm}. The data sets are acquired in an interleaved fashion with and without artificial detuning noise added to the G2 gate as illustrated in \autoref{fig:rbm}(a). The noise is quasi-static, white Gaussian Stark shift noise of $\sigma=20$\,kHz to imitate g-factor variability in a qubit ensemble. For the dressed scheme, the average Clifford gate fidelity and the noise coherence are found to be $98.6(14)\,\%$ and $99.1(14)\,\%$ without added noise, and $95.2(48)\,\%$ and $98.3(48)\,\%$ with added noise, respectively. For the SMART protocol we measure $99.1(9)\,\%$ and $99.4(6)\,\%$ without added noise, and $98.2(18)\,\%$ and $99.0(11)\,\%$ with added noise. The SMART qubit protocol is more robust against detuning noise, dropping by less than $1 \,\%$ in both average Clifford gate fidelity and noise coherence. From the randomised benchmarking data discussed above coherence times can be extracted. These are found to be in the ms range for both the dressed and SMART qubit protocol (see \aref{app:tds}).

The fidelities measured here are most likely limited by the Rabi frequency, which we kept below $0.5 $\,MHz to reduce excitation of two-level charge fluctuators by the stray electric field from the on-chip microwave antenna. To work at higher Rabi frequencies, a microwave antenna or a microwave cavity with lower electric field is required \cite{vahapoglu2021singleelectron,vallabhapurapu2021fast}. The Rabi frequency limits the gate speed in the SMART qubit protocol, since one gate lasts for at least one period of the global field modulation ($\geq T_{\rm{mod}}\propto 1/\Omega_{\rm{R}}$). The duration of a SMART qubit protocol gate compared to a square pulse conventional qubit gate is therefore necessarily longer. The linearity of the Stark shift \cite{chan2018assessment} is another factor affecting the calibration of the SMART gates. From  \autoref{fig:stark} we know that the Stark shift is not perfectly linear, hence it might be necessary to scale the amplitudes making it a more complex process.

\section*{Summary}

We have shown that there exist optimal modulation conditions for a global dressing field to make spin qubits more robust to detuning and microwave amplitude noise while the qubits are individually addressable via electrical Stark shift control. We also suggest more advanced modulation protocols to account for higher order noise. We demonstrate universal control with process tomography as well as randomised benchmarking with fidelities $>99\,\%$. Our experimental results confirm the expected improvement in coherence as compared to a conventionally dressed qubit. The SMART protocol demonstrated here can be implemented with any qubit that allows dressing, providing robustness to qubit variability and improving the prospects for scaling up to full-scale quantum processors.

\section*{Acknowledgements}

We acknowledge support from the Australian Research Council (FL190100167 and CE170100012), the US Army Research Office (W911NF-17-1-0198), and the NSW Node of the Australian National Fabrication Facility. The views and conclusions contained in this document are those of the authors and should not be interpreted as representing the official policies, either expressed or implied, of the Army Research Office or the U.S. Government. The U.S. Government is authorized to reproduce and distribute reprints for Government purposes notwithstanding any copyright notation herein. I.H and A.E.S acknowledge support from Sydney Quantum Academy.

\section*{Author contributions}

I.H. and C.H.Y. performed the experiments and analysed the data. I.H., C.H.Y, A.E.S., A.L. and A.S. discussed the results.  K.W.C. and F.E.H. fabricated the device. K.M.I. prepared and supplied the $^{28}$Si wafer. I.H. and C.H.Y. wrote the
manuscript with input from all co-authors. A.S.D. and C.H.Y. supervised the project.

\appendix

\section{Hamiltonian deduction}
\label{app:Hd}
The laboratory frame Hamiltonian is given by

\begin{equation}
    H_{\rm{lab}}^{\rm{cos}} = \frac{h}{2}\left(\nu(t)\sigma_z+\Omega_{\text{R}}\sqrt{2}\cos{(2\pi{f}_{\rm{mod}}t)}\cos{(2\pi{f}_{\rm{mw}}t)}\sigma_x\right),
    \label{eq:gl2}
\end{equation}
the rotating frame Hamiltonian by
\begin{equation}
    H_{\rm{rot}}^{\rm{cos}} = \frac{h}{2}\left(\Delta\nu(t)\sigma_z+\Omega_{\rm{R}}\sqrt{2}\cos{(2\pi{f}_{\rm{mod}}t)}\sigma_x\right),
    \label{eq:gl3}
\end{equation}
and the rotating frame Hamiltonian using the dressed basis by
\begin{equation}
    H_\rho^{\rm{cos}} = \frac{h}{2}\left(\Omega_{\text{R}}\sqrt{2}\cos(2\pi{f}_{\text{mod}}t)\sigma_z+\Delta\nu(t)\sigma_x\right).
    \label{eq:gl}
\end{equation}
Here, $h$ is Planck's constant, $\nu(t)$ the qubit Larmor frequency, ${f}_{\text{mod}}$ the amplitude modulation frequency, $f_{\rm{mw}}$ the microwave frequency, $\Omega_{\text{R}}$ the Rabi frequency and $\Delta\nu(t)$ the frequency detuning between the Zeeman and the microwave frequency.

\section{Frequency feedback protocol}
\label{app:freq}
Frequency feedback is applied according to Ref. \onlinecite{huang2019fidelity} to compensate for magnetic field decay, residual $^{29}$Si nuclear spins and charge fluctuations causing drifts and jumps of the electron spin resonance frequencies.

\section{Process tomography processing}
\label{app:atomo}
The following procedure is applied when converting the state tomography data into the $\chi$ matrix representation in \autoref{fig:tomo}. After extracting the x-, y- and z probabilities from the six measured projection in the raw data \cite{yang2019silicon}, taking $120$ shots with three repetitions for six different initial states ($\pm{x},\pm{y},\pm{z}$), the density matrix is calculated according to

\begin{equation}
    \rho=\frac{1}{2}\big(\mathbf{I}+P_x\sigma_x+P_y\sigma_y+P_z\sigma_z\big).
\end{equation}

We denote the quantum operation map that maps a quantum state $\rho(0)$ to $\rho(T)$ over time $T$ with a certain control sequence  $\mathcal{E}$
\begin{equation}
    \rho(T)=\mathcal{E}(\rho(0)).
\end{equation}

The following equation is used to find the superoperator $\hat{\mathcal{E}}$, by vectorising the series of $\rho$

\begin{equation}
    \vec{\vec{\rho}}(t) = [\vec{\rho}_1(t),\vec{\rho}_2(t),...,\vec{\rho}_n(t)]
\end{equation}

\begin{equation}
    \hat{\mathcal{E}} = \vec{\vec{\rho}}(T)\vec{\vec{\rho}}(0)^+.
    \label{eq:statetomo}
\end{equation}

Here, we have extended the total control sequence time to $8T$, where we obtained 9 sets of density matrices $\vec{\vec{\rho}}(0), \vec{\vec{\rho}}(T), \vec{\vec{\rho}}(2T),...,\vec{\vec{\rho}}(8T)$. This gives a better fit for the superoperator that has less time dependent variations. Now \autoref{eq:statetomo} can be written as:

\begin{equation}
    \bar{\mathcal{E}} =[\vec{\vec{\rho}}(8T),\vec{\vec{\rho}}(7T),...,\vec{\vec{\rho}}(T)] [\vec{\vec{\rho}}(7T),\vec{\vec{\rho}}(6T),...,\vec{\vec{\rho}}(0)]^+.
\end{equation}
The $\chi$ matrix is then found from
\begin{equation}
    \chi_{ij}=\frac{1}{4}\rm{Tr}\big((\sigma_i\otimes\sigma_j)^\dagger\bar{\mathcal{E}}\big).
\end{equation}
Here $\chi$ completely describes the process in the Pauli basis.

\section{Amplitude correction pulse sequence}
\label{app:amp}
\begin{figure}[htb]
    \centering
    \includegraphics[width=6.5cm]{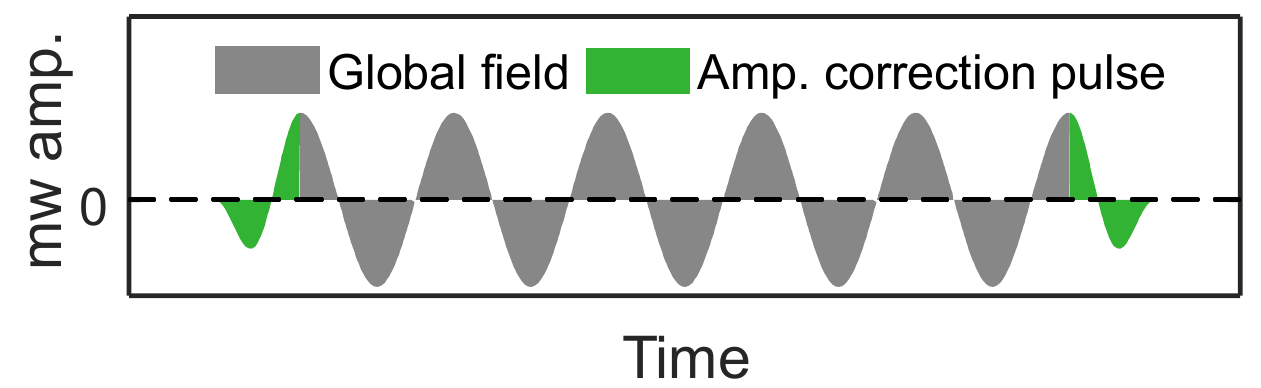}
    \caption{Sequence added before and after the desired microwave modulation in order to avoid jumps in the amplitude, for example making sure a cosine wave has a smooth transition from zero to the amplitude.}
    \label{fig:seq}
\end{figure}

Due to final rise time and power the cosine modulation is sandwiched between a pulse sequence and its mirror symmetric version. The sequence by itself results in identity operation and simultaneously brings the amplitude from zero to the amplitude of the cosine in a smooth manner as shown in \autoref{fig:seq}.

\begin{figure*}[hbt]
    \centering
    \includegraphics[width=\textwidth]{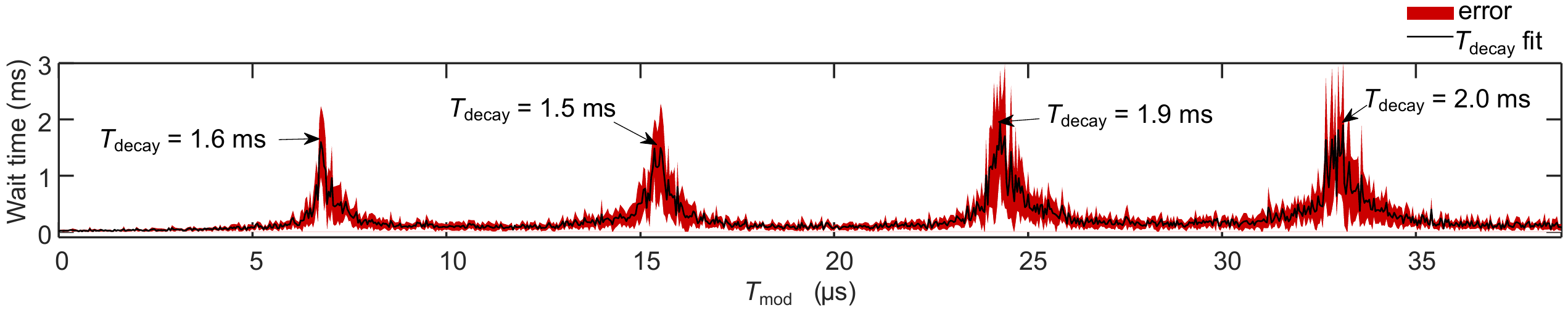}
    \caption{Coherence times extracted from the Ramsey data in \autoref{fig:ramsey}(a) with error bars. The times are extracted with an exponential decay times a cosine term.}
    \label{fig:tds}
\end{figure*}

\section{$T_{\text{decay}}$}
\label{app:tds}
The $T_{\text{decay}}$ times extracted from \autoref{fig:ramsey} is given in \autoref{fig:tds}. We also extract the coherence times corresponding to the randomised benchmarking data in \autoref{fig:rbm}, displayed in \autoref{tab:tab2}.

\begin{table}[h!]
    \centering
     \setlength{\tabcolsep}{0pt}
        \caption{Coherence times extracted from randomised benchmarking data where $\Omega_{\rm R}=0.33\,$MHz, $T_{\rm mod}=5.15$\,\textmu{s} and the gate lengths are $5/\Omega_{\rm R}$ and $3\times T_{\rm mod}$ for the dressed and SMART protocol, respectively.
        }
        \setlength\extrarowheight{2pt}
        \begin{tabular}{c|cc|cc}
        \hline
        \hline
        \multicolumn{1}{c|}{$\hspace{1.4cm}$} &
        \multicolumn{2}{c|}{$\hspace{0.35cm}$Dressed time (\textmu{s})$\hspace{0.35cm}$} & \multicolumn{2}{c}{$\hspace{0.35cm}$SMART (sin) time (\textmu{s})$\hspace{0.35cm}$} \\
        \hline
        \multicolumn{1}{c|}{Noise} &
        \multicolumn{1}{c}{$\hspace{0.25cm}$0\,kHz  $\hspace{0.25cm}$} &
        \multicolumn{1}{c|}{$\hspace{0.25cm}$20\,kHz $\hspace{0.25cm}$} &  
        \multicolumn{1}{c}{$\hspace{0.25cm}$0\,kHz$\hspace{0.25cm}$} &
        \multicolumn{1}{c}{$\hspace{0.25cm}$20\,kHz $\hspace{0.25cm}$}  \\
        \hline
        Overlap & 1097(175) & 364(157)  & 1721(341) &  785(220) \\
        Purity &1785(518) &   1002(290) &  2656(923) & 1534(381) \\
        \hline
        \hline
    \end{tabular}
    \label{tab:tab2}
\end{table}

\bibliography{shaped}

\begin{thebibliography}{36}%
\makeatletter
\providecommand \@ifxundefined [1]{%
 \@ifx{#1\undefined}
}%
\providecommand \@ifnum [1]{%
 \ifnum #1\expandafter \@firstoftwo
 \else \expandafter \@secondoftwo
 \fi
}%
\providecommand \@ifx [1]{%
 \ifx #1\expandafter \@firstoftwo
 \else \expandafter \@secondoftwo
 \fi
}%
\providecommand \natexlab [1]{#1}%
\providecommand \enquote  [1]{``#1''}%
\providecommand \bibnamefont  [1]{#1}%
\providecommand \bibfnamefont [1]{#1}%
\providecommand \citenamefont [1]{#1}%
\providecommand \href@noop [0]{\@secondoftwo}%
\providecommand \href [0]{\begingroup \@sanitize@url \@href}%
\providecommand \@href[1]{\@@startlink{#1}\@@href}%
\providecommand \@@href[1]{\endgroup#1\@@endlink}%
\providecommand \@sanitize@url [0]{\catcode `\\12\catcode `\$12\catcode
  `\&12\catcode `\#12\catcode `\^12\catcode `\_12\catcode `\%12\relax}%
\providecommand \@@startlink[1]{}%
\providecommand \@@endlink[0]{}%
\providecommand \url  [0]{\begingroup\@sanitize@url \@url }%
\providecommand \@url [1]{\endgroup\@href {#1}{\urlprefix }}%
\providecommand \urlprefix  [0]{URL }%
\providecommand \Eprint [0]{\href }%
\providecommand \doibase [0]{http://dx.doi.org/}%
\providecommand \selectlanguage [0]{\@gobble}%
\providecommand \bibinfo  [0]{\@secondoftwo}%
\providecommand \bibfield  [0]{\@secondoftwo}%
\providecommand \translation [1]{[#1]}%
\providecommand \BibitemOpen [0]{}%
\providecommand \bibitemStop [0]{}%
\providecommand \bibitemNoStop [0]{.\EOS\space}%
\providecommand \EOS [0]{\spacefactor3000\relax}%
\providecommand \BibitemShut  [1]{\csname bibitem#1\endcsname}%
\let\auto@bib@innerbib\@empty
%</preamble>
\bibitem [{\citenamefont {Bennett}\ \emph {et~al.}(1996)\citenamefont
  {Bennett}, \citenamefont {DiVincenzo}, \citenamefont {Smolin},\ and\
  \citenamefont {Wootters}}]{bennett1996mixedstate}%
  \BibitemOpen
  \bibfield  {author} {\bibinfo {author} {\bibfnamefont {C.~H.}\ \bibnamefont
  {Bennett}}, \bibinfo {author} {\bibfnamefont {D.~P.}\ \bibnamefont
  {DiVincenzo}}, \bibinfo {author} {\bibfnamefont {J.~A.}\ \bibnamefont
  {Smolin}}, \ and\ \bibinfo {author} {\bibfnamefont {W.~K.}\ \bibnamefont
  {Wootters}},\ }\href {\doibase 10.1103/PhysRevA.54.3824} {\bibfield
  {journal} {\bibinfo  {journal} {Physical Review A}\ }\textbf {\bibinfo
  {volume} {54}},\ \bibinfo {pages} {3824} (\bibinfo {year}
  {1996})}\BibitemShut {NoStop}%
\bibitem [{\citenamefont {Devitt}\ \emph {et~al.}(2013)\citenamefont {Devitt},
  \citenamefont {Munro},\ and\ \citenamefont {Nemoto}}]{devitt2013quantum}%
  \BibitemOpen
  \bibfield  {author} {\bibinfo {author} {\bibfnamefont {S.~J.}\ \bibnamefont
  {Devitt}}, \bibinfo {author} {\bibfnamefont {W.~J.}\ \bibnamefont {Munro}}, \
  and\ \bibinfo {author} {\bibfnamefont {K.}~\bibnamefont {Nemoto}},\ }\href
  {\doibase 10.1088/0034-4885/76/7/076001} {\bibfield  {journal} {\bibinfo
  {journal} {Reports on Progress in Physics}\ }\textbf {\bibinfo {volume}
  {76}},\ \bibinfo {pages} {076001} (\bibinfo {year} {2013})}\BibitemShut
  {NoStop}%
\bibitem [{\citenamefont {Knill}(2005)}]{knill2005quantum}%
  \BibitemOpen
  \bibfield  {author} {\bibinfo {author} {\bibfnamefont {E.}~\bibnamefont
  {Knill}},\ }\href {\doibase 10.1038/nature03350} {\bibfield  {journal}
  {\bibinfo  {journal} {Nature}\ }\textbf {\bibinfo {volume} {434}},\ \bibinfo
  {pages} {39} (\bibinfo {year} {2005})}\BibitemShut {NoStop}%
\bibitem [{\citenamefont {Fowler}\ \emph {et~al.}(2012)\citenamefont {Fowler},
  \citenamefont {Mariantoni}, \citenamefont {Martinis},\ and\ \citenamefont
  {Cleland}}]{fowler2012surface}%
  \BibitemOpen
  \bibfield  {author} {\bibinfo {author} {\bibfnamefont {A.~G.}\ \bibnamefont
  {Fowler}}, \bibinfo {author} {\bibfnamefont {M.}~\bibnamefont {Mariantoni}},
  \bibinfo {author} {\bibfnamefont {J.~M.}\ \bibnamefont {Martinis}}, \ and\
  \bibinfo {author} {\bibfnamefont {A.~N.}\ \bibnamefont {Cleland}},\ }\href
  {\doibase 10.1103/PhysRevA.86.032324} {\bibfield  {journal} {\bibinfo
  {journal} {Physical Review A}\ }\textbf {\bibinfo {volume} {86}},\ \bibinfo
  {pages} {032324} (\bibinfo {year} {2012})}\BibitemShut {NoStop}%
\bibitem [{\citenamefont {Morton}\ \emph {et~al.}(2011)\citenamefont {Morton},
  \citenamefont {McCamey}, \citenamefont {Eriksson},\ and\ \citenamefont
  {Lyon}}]{Morton2011Embracing}%
  \BibitemOpen
  \bibfield  {author} {\bibinfo {author} {\bibfnamefont {J.~J.~L.}\
  \bibnamefont {Morton}}, \bibinfo {author} {\bibfnamefont {D.~R.}\
  \bibnamefont {McCamey}}, \bibinfo {author} {\bibfnamefont {M.~A.}\
  \bibnamefont {Eriksson}}, \ and\ \bibinfo {author} {\bibfnamefont {S.~A.}\
  \bibnamefont {Lyon}},\ }\href {\doibase 10.1038/nature10681} {\bibfield
  {journal} {\bibinfo  {journal} {Nature}\ }\textbf {\bibinfo {volume} {479}},\
  \bibinfo {pages} {345} (\bibinfo {year} {2011})}\BibitemShut {NoStop}%
\bibitem [{\citenamefont {Veldhorst}\ \emph {et~al.}(2014)\citenamefont
  {Veldhorst}, \citenamefont {Hwang}, \citenamefont {Yang}, \citenamefont
  {Leenstra}, \citenamefont {{de Ronde}}, \citenamefont {Dehollain},
  \citenamefont {Muhonen}, \citenamefont {Hudson}, \citenamefont {Itoh},
  \citenamefont {Morello},\ and\ \citenamefont
  {Dzurak}}]{veldhorst2014addressable}%
  \BibitemOpen
  \bibfield  {author} {\bibinfo {author} {\bibfnamefont {M.}~\bibnamefont
  {Veldhorst}}, \bibinfo {author} {\bibfnamefont {J.~C.~C.}\ \bibnamefont
  {Hwang}}, \bibinfo {author} {\bibfnamefont {C.~H.}\ \bibnamefont {Yang}},
  \bibinfo {author} {\bibfnamefont {A.~W.}\ \bibnamefont {Leenstra}}, \bibinfo
  {author} {\bibfnamefont {B.}~\bibnamefont {{de Ronde}}}, \bibinfo {author}
  {\bibfnamefont {J.~P.}\ \bibnamefont {Dehollain}}, \bibinfo {author}
  {\bibfnamefont {J.~T.}\ \bibnamefont {Muhonen}}, \bibinfo {author}
  {\bibfnamefont {F.~E.}\ \bibnamefont {Hudson}}, \bibinfo {author}
  {\bibfnamefont {K.~M.}\ \bibnamefont {Itoh}}, \bibinfo {author}
  {\bibfnamefont {A.}~\bibnamefont {Morello}}, \ and\ \bibinfo {author}
  {\bibfnamefont {A.~S.}\ \bibnamefont {Dzurak}},\ }\href {\doibase
  10.1038/nnano.2014.216} {\bibfield  {journal} {\bibinfo  {journal} {Nature
  Nanotechnology}\ }\textbf {\bibinfo {volume} {9}},\ \bibinfo {pages} {981}
  (\bibinfo {year} {2014})}\BibitemShut {NoStop}%
\bibitem [{\citenamefont {Veldhorst}\ \emph
  {et~al.}(2015{\natexlab{a}})\citenamefont {Veldhorst}, \citenamefont {Yang},
  \citenamefont {Hwang}, \citenamefont {Huang}, \citenamefont {Dehollain},
  \citenamefont {Muhonen}, \citenamefont {Simmons}, \citenamefont {Laucht},
  \citenamefont {Hudson}, \citenamefont {Itoh}, \citenamefont {Morello},\ and\
  \citenamefont {Dzurak}}]{veldhorst2015twoqubit}%
  \BibitemOpen
  \bibfield  {author} {\bibinfo {author} {\bibfnamefont {M.}~\bibnamefont
  {Veldhorst}}, \bibinfo {author} {\bibfnamefont {C.~H.}\ \bibnamefont {Yang}},
  \bibinfo {author} {\bibfnamefont {J.~C.~C.}\ \bibnamefont {Hwang}}, \bibinfo
  {author} {\bibfnamefont {W.}~\bibnamefont {Huang}}, \bibinfo {author}
  {\bibfnamefont {J.~P.}\ \bibnamefont {Dehollain}}, \bibinfo {author}
  {\bibfnamefont {J.~T.}\ \bibnamefont {Muhonen}}, \bibinfo {author}
  {\bibfnamefont {S.}~\bibnamefont {Simmons}}, \bibinfo {author} {\bibfnamefont
  {A.}~\bibnamefont {Laucht}}, \bibinfo {author} {\bibfnamefont {F.~E.}\
  \bibnamefont {Hudson}}, \bibinfo {author} {\bibfnamefont {K.~M.}\
  \bibnamefont {Itoh}}, \bibinfo {author} {\bibfnamefont {A.}~\bibnamefont
  {Morello}}, \ and\ \bibinfo {author} {\bibfnamefont {A.~S.}\ \bibnamefont
  {Dzurak}},\ }\href {\doibase 10.1038/nature15263} {\bibfield  {journal}
  {\bibinfo  {journal} {Nature}\ }\textbf {\bibinfo {volume} {526}},\ \bibinfo
  {pages} {410} (\bibinfo {year} {2015}{\natexlab{a}})}\BibitemShut {NoStop}%
\bibitem [{\citenamefont {Veldhorst}\ \emph {et~al.}(2017)\citenamefont
  {Veldhorst}, \citenamefont {Eenink}, \citenamefont {Yang},\ and\
  \citenamefont {Dzurak}}]{veldhorst2017silicon}%
  \BibitemOpen
  \bibfield  {author} {\bibinfo {author} {\bibfnamefont {M.}~\bibnamefont
  {Veldhorst}}, \bibinfo {author} {\bibfnamefont {H.~G.~J.}\ \bibnamefont
  {Eenink}}, \bibinfo {author} {\bibfnamefont {C.~H.}\ \bibnamefont {Yang}}, \
  and\ \bibinfo {author} {\bibfnamefont {A.~S.}\ \bibnamefont {Dzurak}},\
  }\href {\doibase 10.1038/s41467-017-01905-6} {\bibfield  {journal} {\bibinfo
  {journal} {Nature Communications}\ }\textbf {\bibinfo {volume} {8}},\
  \bibinfo {pages} {1766} (\bibinfo {year} {2017})}\BibitemShut {NoStop}%
\bibitem [{\citenamefont {Zhang}\ \emph {et~al.}(2018)\citenamefont {Zhang},
  \citenamefont {Li}, \citenamefont {Cao}, \citenamefont {Xiao}, \citenamefont
  {Guo},\ and\ \citenamefont {Guo}}]{Zhang2018Semiconductor}%
  \BibitemOpen
  \bibfield  {author} {\bibinfo {author} {\bibfnamefont {X.}~\bibnamefont
  {Zhang}}, \bibinfo {author} {\bibfnamefont {H.-O.}\ \bibnamefont {Li}},
  \bibinfo {author} {\bibfnamefont {G.}~\bibnamefont {Cao}}, \bibinfo {author}
  {\bibfnamefont {M.}~\bibnamefont {Xiao}}, \bibinfo {author} {\bibfnamefont
  {G.-C.}\ \bibnamefont {Guo}}, \ and\ \bibinfo {author} {\bibfnamefont
  {G.-P.}\ \bibnamefont {Guo}},\ }\href {\doibase 10.1093/nsr/nwy153}
  {\bibfield  {journal} {\bibinfo  {journal} {National Science Review}\
  }\textbf {\bibinfo {volume} {6}},\ \bibinfo {pages} {32} (\bibinfo {year}
  {2018})}\BibitemShut {NoStop}%
\bibitem [{\citenamefont {Hensen}\ \emph {et~al.}(2020)\citenamefont {Hensen},
  \citenamefont {Wei~Huang}, \citenamefont {Yang}, \citenamefont {Chan},
  \citenamefont {Yoneda}, \citenamefont {Tanttu}, \citenamefont {Hudson},
  \citenamefont {Laucht}, \citenamefont {Itoh}, \citenamefont {Ladd},
  \citenamefont {Morello},\ and\ \citenamefont {Dzurak}}]{hensen2020silicon}%
  \BibitemOpen
  \bibfield  {author} {\bibinfo {author} {\bibfnamefont {B.}~\bibnamefont
  {Hensen}}, \bibinfo {author} {\bibfnamefont {W.}~\bibnamefont {Wei~Huang}},
  \bibinfo {author} {\bibfnamefont {C.-H.}\ \bibnamefont {Yang}}, \bibinfo
  {author} {\bibfnamefont {K.~W.}\ \bibnamefont {Chan}}, \bibinfo {author}
  {\bibfnamefont {J.}~\bibnamefont {Yoneda}}, \bibinfo {author} {\bibfnamefont
  {T.}~\bibnamefont {Tanttu}}, \bibinfo {author} {\bibfnamefont {F.~E.}\
  \bibnamefont {Hudson}}, \bibinfo {author} {\bibfnamefont {A.}~\bibnamefont
  {Laucht}}, \bibinfo {author} {\bibfnamefont {K.~M.}\ \bibnamefont {Itoh}},
  \bibinfo {author} {\bibfnamefont {T.~D.}\ \bibnamefont {Ladd}}, \bibinfo
  {author} {\bibfnamefont {A.}~\bibnamefont {Morello}}, \ and\ \bibinfo
  {author} {\bibfnamefont {A.~S.}\ \bibnamefont {Dzurak}},\ }\href {\doibase
  10.1038/s41565-019-0587-7} {\bibfield  {journal} {\bibinfo  {journal} {Nature
  Nanotechnology}\ }\textbf {\bibinfo {volume} {15}},\ \bibinfo {pages} {13}
  (\bibinfo {year} {2020})}\BibitemShut {NoStop}%
\bibitem [{\citenamefont {Zhao}\ \emph {et~al.}(2019)\citenamefont {Zhao},
  \citenamefont {Tanttu}, \citenamefont {Tan}, \citenamefont {Hensen},
  \citenamefont {Chan}, \citenamefont {Hwang}, \citenamefont {Leon},
  \citenamefont {Yang}, \citenamefont {Gilbert}, \citenamefont {Hudson},
  \citenamefont {Itoh}, \citenamefont {Kiselev}, \citenamefont {Ladd},
  \citenamefont {Morello}, \citenamefont {Laucht},\ and\ \citenamefont
  {Dzurak}}]{zhao2019singlespin}%
  \BibitemOpen
  \bibfield  {author} {\bibinfo {author} {\bibfnamefont {R.}~\bibnamefont
  {Zhao}}, \bibinfo {author} {\bibfnamefont {T.}~\bibnamefont {Tanttu}},
  \bibinfo {author} {\bibfnamefont {K.~Y.}\ \bibnamefont {Tan}}, \bibinfo
  {author} {\bibfnamefont {B.}~\bibnamefont {Hensen}}, \bibinfo {author}
  {\bibfnamefont {K.~W.}\ \bibnamefont {Chan}}, \bibinfo {author}
  {\bibfnamefont {J.~C.~C.}\ \bibnamefont {Hwang}}, \bibinfo {author}
  {\bibfnamefont {R.~C.~C.}\ \bibnamefont {Leon}}, \bibinfo {author}
  {\bibfnamefont {C.~H.}\ \bibnamefont {Yang}}, \bibinfo {author}
  {\bibfnamefont {W.}~\bibnamefont {Gilbert}}, \bibinfo {author} {\bibfnamefont
  {F.~E.}\ \bibnamefont {Hudson}}, \bibinfo {author} {\bibfnamefont {K.~M.}\
  \bibnamefont {Itoh}}, \bibinfo {author} {\bibfnamefont {A.~A.}\ \bibnamefont
  {Kiselev}}, \bibinfo {author} {\bibfnamefont {T.~D.}\ \bibnamefont {Ladd}},
  \bibinfo {author} {\bibfnamefont {A.}~\bibnamefont {Morello}}, \bibinfo
  {author} {\bibfnamefont {A.}~\bibnamefont {Laucht}}, \ and\ \bibinfo {author}
  {\bibfnamefont {A.~S.}\ \bibnamefont {Dzurak}},\ }\href {\doibase
  10.1038/s41467-019-13416-7} {\bibfield  {journal} {\bibinfo  {journal}
  {Nature Communications}\ }\textbf {\bibinfo {volume} {10}},\ \bibinfo {pages}
  {5500} (\bibinfo {year} {2019})}\BibitemShut {NoStop}%
\bibitem [{\citenamefont {Veldhorst}\ \emph
  {et~al.}(2015{\natexlab{b}})\citenamefont {Veldhorst}, \citenamefont
  {Ruskov}, \citenamefont {Yang}, \citenamefont {Hwang}, \citenamefont
  {Hudson}, \citenamefont {Flatt{\'e}}, \citenamefont {Tahan}, \citenamefont
  {Itoh}, \citenamefont {Morello},\ and\ \citenamefont
  {Dzurak}}]{veldhorst2015spinorbit}%
  \BibitemOpen
  \bibfield  {author} {\bibinfo {author} {\bibfnamefont {M.}~\bibnamefont
  {Veldhorst}}, \bibinfo {author} {\bibfnamefont {R.}~\bibnamefont {Ruskov}},
  \bibinfo {author} {\bibfnamefont {C.~H.}\ \bibnamefont {Yang}}, \bibinfo
  {author} {\bibfnamefont {J.~C.~C.}\ \bibnamefont {Hwang}}, \bibinfo {author}
  {\bibfnamefont {F.~E.}\ \bibnamefont {Hudson}}, \bibinfo {author}
  {\bibfnamefont {M.~E.}\ \bibnamefont {Flatt{\'e}}}, \bibinfo {author}
  {\bibfnamefont {C.}~\bibnamefont {Tahan}}, \bibinfo {author} {\bibfnamefont
  {K.~M.}\ \bibnamefont {Itoh}}, \bibinfo {author} {\bibfnamefont
  {A.}~\bibnamefont {Morello}}, \ and\ \bibinfo {author} {\bibfnamefont
  {A.~S.}\ \bibnamefont {Dzurak}},\ }\href {\doibase
  10.1103/PhysRevB.92.201401} {\bibfield  {journal} {\bibinfo  {journal}
  {Physical Review B}\ }\textbf {\bibinfo {volume} {92}},\ \bibinfo {pages}
  {201401} (\bibinfo {year} {2015}{\natexlab{b}})}\BibitemShut {NoStop}%
\bibitem [{\citenamefont {Ruskov}\ \emph {et~al.}(2018)\citenamefont {Ruskov},
  \citenamefont {Veldhorst}, \citenamefont {Dzurak},\ and\ \citenamefont
  {Tahan}}]{ruskov2018electron}%
  \BibitemOpen
  \bibfield  {author} {\bibinfo {author} {\bibfnamefont {R.}~\bibnamefont
  {Ruskov}}, \bibinfo {author} {\bibfnamefont {M.}~\bibnamefont {Veldhorst}},
  \bibinfo {author} {\bibfnamefont {A.~S.}\ \bibnamefont {Dzurak}}, \ and\
  \bibinfo {author} {\bibfnamefont {C.}~\bibnamefont {Tahan}},\ }\href
  {\doibase 10.1103/PhysRevB.98.245424} {\bibfield  {journal} {\bibinfo
  {journal} {Physical Review B}\ }\textbf {\bibinfo {volume} {98}},\ \bibinfo
  {pages} {245424} (\bibinfo {year} {2018})}\BibitemShut {NoStop}%
\bibitem [{\citenamefont {Vahapoglu}\ \emph
  {et~al.}(2021{\natexlab{a}})\citenamefont {Vahapoglu}, \citenamefont
  {Slack-Smith}, \citenamefont {Leon}, \citenamefont {Lim}, \citenamefont
  {Hudson}, \citenamefont {Day}, \citenamefont {Tanttu}, \citenamefont {Yang},
  \citenamefont {Laucht}, \citenamefont {Dzurak},\ and\ \citenamefont
  {Pla}}]{vahapoglu2021singleelectron}%
  \BibitemOpen
  \bibfield  {author} {\bibinfo {author} {\bibfnamefont {E.}~\bibnamefont
  {Vahapoglu}}, \bibinfo {author} {\bibfnamefont {J.~P.}\ \bibnamefont
  {Slack-Smith}}, \bibinfo {author} {\bibfnamefont {R.~C.}\ \bibnamefont
  {Leon}}, \bibinfo {author} {\bibfnamefont {W.~H.}\ \bibnamefont {Lim}},
  \bibinfo {author} {\bibfnamefont {F.~E.}\ \bibnamefont {Hudson}}, \bibinfo
  {author} {\bibfnamefont {T.}~\bibnamefont {Day}}, \bibinfo {author}
  {\bibfnamefont {T.}~\bibnamefont {Tanttu}}, \bibinfo {author} {\bibfnamefont
  {C.~H.}\ \bibnamefont {Yang}}, \bibinfo {author} {\bibfnamefont
  {A.}~\bibnamefont {Laucht}}, \bibinfo {author} {\bibfnamefont {A.~S.}\
  \bibnamefont {Dzurak}}, \ and\ \bibinfo {author} {\bibfnamefont {J.~J.}\
  \bibnamefont {Pla}},\ }\href {\doibase 10.1126/sciadv.abg9158} {\bibfield
  {journal} {\bibinfo  {journal} {Science Advances}\ }\textbf {\bibinfo
  {volume} {7}} (\bibinfo {year} {2021}{\natexlab{a}}),\
  10.1126/sciadv.abg9158}\BibitemShut {NoStop}%
\bibitem [{\citenamefont {Vallabhapurapu}\ \emph {et~al.}(2021)\citenamefont
  {Vallabhapurapu}, \citenamefont {{Slack-Smith}}, \citenamefont {Sewani},
  \citenamefont {Adambukulam}, \citenamefont {Morello}, \citenamefont {Pla},\
  and\ \citenamefont {Laucht}}]{vallabhapurapu2021fast}%
  \BibitemOpen
  \bibfield  {author} {\bibinfo {author} {\bibfnamefont {H.~H.}\ \bibnamefont
  {Vallabhapurapu}}, \bibinfo {author} {\bibfnamefont {J.~P.}\ \bibnamefont
  {{Slack-Smith}}}, \bibinfo {author} {\bibfnamefont {V.~K.}\ \bibnamefont
  {Sewani}}, \bibinfo {author} {\bibfnamefont {C.}~\bibnamefont {Adambukulam}},
  \bibinfo {author} {\bibfnamefont {A.}~\bibnamefont {Morello}}, \bibinfo
  {author} {\bibfnamefont {J.~J.}\ \bibnamefont {Pla}}, \ and\ \bibinfo
  {author} {\bibfnamefont {A.}~\bibnamefont {Laucht}},\ }\href@noop {}
  {\bibfield  {journal} {\bibinfo  {journal} {arXiv:2105.06781 [cond-mat,
  physics:quant-ph]}\ } (\bibinfo {year} {2021})},\ \Eprint
  {http://arxiv.org/abs/2105.06781} {arXiv:2105.06781 [cond-mat,
  physics:quant-ph]} \BibitemShut {NoStop}%
\bibitem [{\citenamefont {Seedhouse}\ \emph {et~al.}(2021)\citenamefont
  {Seedhouse}, \citenamefont {Hansen}, \citenamefont {Laucht}, \citenamefont
  {Yang}, \citenamefont {Dzurak},\ and\ \citenamefont
  {Saraiva}}]{seedhouse2021quantum}%
  \BibitemOpen
  \bibfield  {author} {\bibinfo {author} {\bibfnamefont {A.~E.}\ \bibnamefont
  {Seedhouse}}, \bibinfo {author} {\bibfnamefont {I.}~\bibnamefont {Hansen}},
  \bibinfo {author} {\bibfnamefont {A.}~\bibnamefont {Laucht}}, \bibinfo
  {author} {\bibfnamefont {C.~H.}\ \bibnamefont {Yang}}, \bibinfo {author}
  {\bibfnamefont {A.~S.}\ \bibnamefont {Dzurak}}, \ and\ \bibinfo {author}
  {\bibfnamefont {A.}~\bibnamefont {Saraiva}},\ }\href@noop {} {\bibfield
  {journal} {\bibinfo  {journal} {arXiv:2108.00798 [cond-mat,
  physics:quant-ph]}\ } (\bibinfo {year} {2021})},\ \Eprint
  {http://arxiv.org/abs/2108.00798} {arXiv:2108.00798 [cond-mat,
  physics:quant-ph]} \BibitemShut {NoStop}%
\bibitem [{\citenamefont {Vahapoglu}\ \emph
  {et~al.}(2021{\natexlab{b}})\citenamefont {Vahapoglu}, \citenamefont
  {{Slack-Smith}}, \citenamefont {Leon}, \citenamefont {Lim}, \citenamefont
  {Hudson}, \citenamefont {Day}, \citenamefont {Cifuentes}, \citenamefont
  {Tanttu}, \citenamefont {Yang}, \citenamefont {Saraiva}, \citenamefont
  {Thewalt}, \citenamefont {Laucht}, \citenamefont {Dzurak},\ and\
  \citenamefont {Pla}}]{vahapoglu2021coherent}%
  \BibitemOpen
  \bibfield  {author} {\bibinfo {author} {\bibfnamefont {E.}~\bibnamefont
  {Vahapoglu}}, \bibinfo {author} {\bibfnamefont {J.~P.}\ \bibnamefont
  {{Slack-Smith}}}, \bibinfo {author} {\bibfnamefont {R.~C.~C.}\ \bibnamefont
  {Leon}}, \bibinfo {author} {\bibfnamefont {W.~H.}\ \bibnamefont {Lim}},
  \bibinfo {author} {\bibfnamefont {F.~E.}\ \bibnamefont {Hudson}}, \bibinfo
  {author} {\bibfnamefont {T.}~\bibnamefont {Day}}, \bibinfo {author}
  {\bibfnamefont {J.~D.}\ \bibnamefont {Cifuentes}}, \bibinfo {author}
  {\bibfnamefont {T.}~\bibnamefont {Tanttu}}, \bibinfo {author} {\bibfnamefont
  {C.~H.}\ \bibnamefont {Yang}}, \bibinfo {author} {\bibfnamefont
  {A.}~\bibnamefont {Saraiva}}, \bibinfo {author} {\bibfnamefont {M.~L.~W.}\
  \bibnamefont {Thewalt}}, \bibinfo {author} {\bibfnamefont {A.}~\bibnamefont
  {Laucht}}, \bibinfo {author} {\bibfnamefont {A.~S.}\ \bibnamefont {Dzurak}},
  \ and\ \bibinfo {author} {\bibfnamefont {J.~J.}\ \bibnamefont {Pla}},\
  }\href@noop {} {\bibfield  {journal} {\bibinfo  {journal} {arXiv:2107.14622
  [cond-mat]}\ } (\bibinfo {year} {2021}{\natexlab{b}})},\ \Eprint
  {http://arxiv.org/abs/2107.14622} {arXiv:2107.14622 [cond-mat]} \BibitemShut
  {NoStop}%
\bibitem [{\citenamefont {Kane}(1998)}]{kane1998siliconbased}%
  \BibitemOpen
  \bibfield  {author} {\bibinfo {author} {\bibfnamefont {B.~E.}\ \bibnamefont
  {Kane}},\ }\href {\doibase 10.1038/30156} {\bibfield  {journal} {\bibinfo
  {journal} {Nature}\ }\textbf {\bibinfo {volume} {393}},\ \bibinfo {pages}
  {133} (\bibinfo {year} {1998})}\BibitemShut {NoStop}%
\bibitem [{\citenamefont {Laucht}\ \emph {et~al.}(2015)\citenamefont {Laucht},
  \citenamefont {Muhonen}, \citenamefont {Mohiyaddin}, \citenamefont {Kalra},
  \citenamefont {Dehollain}, \citenamefont {Freer}, \citenamefont {Hudson},
  \citenamefont {Veldhorst}, \citenamefont {Rahman}, \citenamefont {Klimeck},
  \citenamefont {Itoh}, \citenamefont {Jamieson}, \citenamefont {McCallum},
  \citenamefont {Dzurak},\ and\ \citenamefont
  {Morello}}]{laucht2015electrically}%
  \BibitemOpen
  \bibfield  {author} {\bibinfo {author} {\bibfnamefont {A.}~\bibnamefont
  {Laucht}}, \bibinfo {author} {\bibfnamefont {J.~T.}\ \bibnamefont {Muhonen}},
  \bibinfo {author} {\bibfnamefont {F.~A.}\ \bibnamefont {Mohiyaddin}},
  \bibinfo {author} {\bibfnamefont {R.}~\bibnamefont {Kalra}}, \bibinfo
  {author} {\bibfnamefont {J.~P.}\ \bibnamefont {Dehollain}}, \bibinfo {author}
  {\bibfnamefont {S.}~\bibnamefont {Freer}}, \bibinfo {author} {\bibfnamefont
  {F.~E.}\ \bibnamefont {Hudson}}, \bibinfo {author} {\bibfnamefont
  {M.}~\bibnamefont {Veldhorst}}, \bibinfo {author} {\bibfnamefont
  {R.}~\bibnamefont {Rahman}}, \bibinfo {author} {\bibfnamefont
  {G.}~\bibnamefont {Klimeck}}, \bibinfo {author} {\bibfnamefont {K.~M.}\
  \bibnamefont {Itoh}}, \bibinfo {author} {\bibfnamefont {D.~N.}\ \bibnamefont
  {Jamieson}}, \bibinfo {author} {\bibfnamefont {J.~C.}\ \bibnamefont
  {McCallum}}, \bibinfo {author} {\bibfnamefont {A.~S.}\ \bibnamefont
  {Dzurak}}, \ and\ \bibinfo {author} {\bibfnamefont {A.}~\bibnamefont
  {Morello}},\ }\href {\doibase 10.1126/sciadv.1500022} {\bibfield  {journal}
  {\bibinfo  {journal} {Science Advances}\ }\textbf {\bibinfo {volume} {1}},\
  \bibinfo {pages} {e1500022} (\bibinfo {year} {2015})}\BibitemShut {NoStop}%
\bibitem [{\citenamefont {Golter}\ \emph {et~al.}(2014)\citenamefont {Golter},
  \citenamefont {Baldwin},\ and\ \citenamefont {Wang}}]{golter2014protecting}%
  \BibitemOpen
  \bibfield  {author} {\bibinfo {author} {\bibfnamefont {D.~A.}\ \bibnamefont
  {Golter}}, \bibinfo {author} {\bibfnamefont {T.~K.}\ \bibnamefont {Baldwin}},
  \ and\ \bibinfo {author} {\bibfnamefont {H.}~\bibnamefont {Wang}},\ }\href
  {\doibase 10.1103/PhysRevLett.113.237601} {\bibfield  {journal} {\bibinfo
  {journal} {Physical Review Letters}\ }\textbf {\bibinfo {volume} {113}},\
  \bibinfo {pages} {237601} (\bibinfo {year} {2014})}\BibitemShut {NoStop}%
\bibitem [{\citenamefont {Wu}\ \emph {et~al.}(2019)\citenamefont {Wu},
  \citenamefont {Amezcua},\ and\ \citenamefont {Wang}}]{wu2019adiabatic}%
  \BibitemOpen
  \bibfield  {author} {\bibinfo {author} {\bibfnamefont {S.-H.}\ \bibnamefont
  {Wu}}, \bibinfo {author} {\bibfnamefont {M.}~\bibnamefont {Amezcua}}, \ and\
  \bibinfo {author} {\bibfnamefont {H.}~\bibnamefont {Wang}},\ }\href {\doibase
  10.1103/PhysRevA.99.063812} {\bibfield  {journal} {\bibinfo  {journal}
  {Physical Review A}\ }\textbf {\bibinfo {volume} {99}},\ \bibinfo {pages}
  {063812} (\bibinfo {year} {2019})}\BibitemShut {NoStop}%
\bibitem [{\citenamefont {Mikelsons}\ \emph {et~al.}(2015)\citenamefont
  {Mikelsons}, \citenamefont {Cohen}, \citenamefont {Retzker},\ and\
  \citenamefont {Plenio}}]{mikelsons2015universal}%
  \BibitemOpen
  \bibfield  {author} {\bibinfo {author} {\bibfnamefont {G.}~\bibnamefont
  {Mikelsons}}, \bibinfo {author} {\bibfnamefont {I.}~\bibnamefont {Cohen}},
  \bibinfo {author} {\bibfnamefont {A.}~\bibnamefont {Retzker}}, \ and\
  \bibinfo {author} {\bibfnamefont {M.~B.}\ \bibnamefont {Plenio}},\ }\href
  {\doibase 10.1088/1367-2630/17/5/053032} {\bibfield  {journal} {\bibinfo
  {journal} {New Journal of Physics}\ }\textbf {\bibinfo {volume} {17}},\
  \bibinfo {pages} {053032} (\bibinfo {year} {2015})}\BibitemShut {NoStop}%
\bibitem [{\citenamefont {Laucht}\ \emph {et~al.}(2017)\citenamefont {Laucht},
  \citenamefont {Kalra}, \citenamefont {Simmons}, \citenamefont {Dehollain},
  \citenamefont {Muhonen}, \citenamefont {Mohiyaddin}, \citenamefont {Freer},
  \citenamefont {Hudson}, \citenamefont {Itoh}, \citenamefont {Jamieson},
  \citenamefont {McCallum}, \citenamefont {Dzurak},\ and\ \citenamefont
  {Morello}}]{laucht2017dressed}%
  \BibitemOpen
  \bibfield  {author} {\bibinfo {author} {\bibfnamefont {A.}~\bibnamefont
  {Laucht}}, \bibinfo {author} {\bibfnamefont {R.}~\bibnamefont {Kalra}},
  \bibinfo {author} {\bibfnamefont {S.}~\bibnamefont {Simmons}}, \bibinfo
  {author} {\bibfnamefont {J.~P.}\ \bibnamefont {Dehollain}}, \bibinfo {author}
  {\bibfnamefont {J.~T.}\ \bibnamefont {Muhonen}}, \bibinfo {author}
  {\bibfnamefont {F.~A.}\ \bibnamefont {Mohiyaddin}}, \bibinfo {author}
  {\bibfnamefont {S.}~\bibnamefont {Freer}}, \bibinfo {author} {\bibfnamefont
  {F.~E.}\ \bibnamefont {Hudson}}, \bibinfo {author} {\bibfnamefont {K.~M.}\
  \bibnamefont {Itoh}}, \bibinfo {author} {\bibfnamefont {D.~N.}\ \bibnamefont
  {Jamieson}}, \bibinfo {author} {\bibfnamefont {J.~C.}\ \bibnamefont
  {McCallum}}, \bibinfo {author} {\bibfnamefont {A.~S.}\ \bibnamefont
  {Dzurak}}, \ and\ \bibinfo {author} {\bibfnamefont {A.}~\bibnamefont
  {Morello}},\ }\href {\doibase 10.1038/nnano.2016.178} {\bibfield  {journal}
  {\bibinfo  {journal} {Nature Nanotechnology}\ }\textbf {\bibinfo {volume}
  {12}},\ \bibinfo {pages} {61} (\bibinfo {year} {2017})}\BibitemShut {NoStop}%
\bibitem [{\citenamefont {Miao}\ \emph {et~al.}(2020)\citenamefont {Miao},
  \citenamefont {Blanton}, \citenamefont {Anderson}, \citenamefont {Bourassa},
  \citenamefont {Crook}, \citenamefont {Wolfowicz}, \citenamefont {Abe},
  \citenamefont {Ohshima},\ and\ \citenamefont
  {Awschalom}}]{Miao2020Universal}%
  \BibitemOpen
  \bibfield  {author} {\bibinfo {author} {\bibfnamefont {K.~C.}\ \bibnamefont
  {Miao}}, \bibinfo {author} {\bibfnamefont {J.~P.}\ \bibnamefont {Blanton}},
  \bibinfo {author} {\bibfnamefont {C.~P.}\ \bibnamefont {Anderson}}, \bibinfo
  {author} {\bibfnamefont {A.}~\bibnamefont {Bourassa}}, \bibinfo {author}
  {\bibfnamefont {A.~L.}\ \bibnamefont {Crook}}, \bibinfo {author}
  {\bibfnamefont {G.}~\bibnamefont {Wolfowicz}}, \bibinfo {author}
  {\bibfnamefont {H.}~\bibnamefont {Abe}}, \bibinfo {author} {\bibfnamefont
  {T.}~\bibnamefont {Ohshima}}, \ and\ \bibinfo {author} {\bibfnamefont
  {D.~D.}\ \bibnamefont {Awschalom}},\ }\href {\doibase
  10.1126/science.abc5186} {\bibfield  {journal} {\bibinfo  {journal}
  {Science}\ } (\bibinfo {year} {2020}),\ 10.1126/science.abc5186}\BibitemShut
  {NoStop}%
\bibitem [{\citenamefont {Jones}\ \emph {et~al.}(2018)\citenamefont {Jones},
  \citenamefont {Fogarty}, \citenamefont {Morello}, \citenamefont {Gyure},
  \citenamefont {Dzurak},\ and\ \citenamefont {Ladd}}]{jones2018logical}%
  \BibitemOpen
  \bibfield  {author} {\bibinfo {author} {\bibfnamefont {C.}~\bibnamefont
  {Jones}}, \bibinfo {author} {\bibfnamefont {M.~A.}\ \bibnamefont {Fogarty}},
  \bibinfo {author} {\bibfnamefont {A.}~\bibnamefont {Morello}}, \bibinfo
  {author} {\bibfnamefont {M.~F.}\ \bibnamefont {Gyure}}, \bibinfo {author}
  {\bibfnamefont {A.~S.}\ \bibnamefont {Dzurak}}, \ and\ \bibinfo {author}
  {\bibfnamefont {T.~D.}\ \bibnamefont {Ladd}},\ }\href {\doibase
  10.1103/PhysRevX.8.021058} {\bibfield  {journal} {\bibinfo  {journal}
  {Physical Review X}\ }\textbf {\bibinfo {volume} {8}},\ \bibinfo {pages}
  {021058} (\bibinfo {year} {2018})}\BibitemShut {NoStop}%
\bibitem [{\citenamefont {Hansen}\ \emph {et~al.}(2021)\citenamefont {Hansen},
  \citenamefont {Seedhouse}, \citenamefont {Saraiva}, \citenamefont {Laucht},
  \citenamefont {Dzurak},\ and\ \citenamefont {Yang}}]{hansen2021smart}%
  \BibitemOpen
  \bibfield  {author} {\bibinfo {author} {\bibfnamefont {I.}~\bibnamefont
  {Hansen}}, \bibinfo {author} {\bibfnamefont {A.~E.}\ \bibnamefont
  {Seedhouse}}, \bibinfo {author} {\bibfnamefont {A.}~\bibnamefont {Saraiva}},
  \bibinfo {author} {\bibfnamefont {A.}~\bibnamefont {Laucht}}, \bibinfo
  {author} {\bibfnamefont {A.~S.}\ \bibnamefont {Dzurak}}, \ and\ \bibinfo
  {author} {\bibfnamefont {C.~H.}\ \bibnamefont {Yang}},\ }\href@noop {}
  {\bibfield  {journal} {\bibinfo  {journal} {arXiv:2108.00776 [cond-mat,
  physics:quant-ph]}\ } (\bibinfo {year} {2021})},\ \Eprint
  {http://arxiv.org/abs/2108.00776} {arXiv:2108.00776 [cond-mat,
  physics:quant-ph]} \BibitemShut {NoStop}%
\bibitem [{\citenamefont {Yoneda}\ \emph {et~al.}(2021)\citenamefont {Yoneda},
  \citenamefont {Huang}, \citenamefont {Feng}, \citenamefont {Yang},
  \citenamefont {Chan}, \citenamefont {Tanttu}, \citenamefont {Gilbert},
  \citenamefont {Leon}, \citenamefont {Hudson}, \citenamefont {Itoh},
  \citenamefont {Morello}, \citenamefont {Bartlett}, \citenamefont {Laucht},
  \citenamefont {Saraiva},\ and\ \citenamefont {Dzurak}}]{yoneda2021coherent}%
  \BibitemOpen
  \bibfield  {author} {\bibinfo {author} {\bibfnamefont {J.}~\bibnamefont
  {Yoneda}}, \bibinfo {author} {\bibfnamefont {W.}~\bibnamefont {Huang}},
  \bibinfo {author} {\bibfnamefont {M.}~\bibnamefont {Feng}}, \bibinfo {author}
  {\bibfnamefont {C.~H.}\ \bibnamefont {Yang}}, \bibinfo {author}
  {\bibfnamefont {K.~W.}\ \bibnamefont {Chan}}, \bibinfo {author}
  {\bibfnamefont {T.}~\bibnamefont {Tanttu}}, \bibinfo {author} {\bibfnamefont
  {W.}~\bibnamefont {Gilbert}}, \bibinfo {author} {\bibfnamefont {R.~C.~C.}\
  \bibnamefont {Leon}}, \bibinfo {author} {\bibfnamefont {F.~E.}\ \bibnamefont
  {Hudson}}, \bibinfo {author} {\bibfnamefont {K.~M.}\ \bibnamefont {Itoh}},
  \bibinfo {author} {\bibfnamefont {A.}~\bibnamefont {Morello}}, \bibinfo
  {author} {\bibfnamefont {S.~D.}\ \bibnamefont {Bartlett}}, \bibinfo {author}
  {\bibfnamefont {A.}~\bibnamefont {Laucht}}, \bibinfo {author} {\bibfnamefont
  {A.}~\bibnamefont {Saraiva}}, \ and\ \bibinfo {author} {\bibfnamefont
  {A.~S.}\ \bibnamefont {Dzurak}},\ }\href {\doibase
  10.1038/s41467-021-24371-7} {\bibfield  {journal} {\bibinfo  {journal}
  {Nature Communications}\ }\textbf {\bibinfo {volume} {12}},\ \bibinfo {pages}
  {4114} (\bibinfo {year} {2021})}\BibitemShut {NoStop}%
\bibitem [{\citenamefont {Yang}\ \emph {et~al.}(2019)\citenamefont {Yang},
  \citenamefont {Chan}, \citenamefont {Harper}, \citenamefont {Huang},
  \citenamefont {Evans}, \citenamefont {Hwang}, \citenamefont {Hensen},
  \citenamefont {Laucht}, \citenamefont {Tanttu}, \citenamefont {Hudson},
  \citenamefont {Flammia}, \citenamefont {Itoh}, \citenamefont {Morello},
  \citenamefont {Bartlett},\ and\ \citenamefont {Dzurak}}]{yang2019silicon}%
  \BibitemOpen
  \bibfield  {author} {\bibinfo {author} {\bibfnamefont {C.~H.}\ \bibnamefont
  {Yang}}, \bibinfo {author} {\bibfnamefont {K.~W.}\ \bibnamefont {Chan}},
  \bibinfo {author} {\bibfnamefont {R.}~\bibnamefont {Harper}}, \bibinfo
  {author} {\bibfnamefont {W.}~\bibnamefont {Huang}}, \bibinfo {author}
  {\bibfnamefont {T.}~\bibnamefont {Evans}}, \bibinfo {author} {\bibfnamefont
  {J.~C.~C.}\ \bibnamefont {Hwang}}, \bibinfo {author} {\bibfnamefont
  {B.}~\bibnamefont {Hensen}}, \bibinfo {author} {\bibfnamefont
  {A.}~\bibnamefont {Laucht}}, \bibinfo {author} {\bibfnamefont
  {T.}~\bibnamefont {Tanttu}}, \bibinfo {author} {\bibfnamefont {F.~E.}\
  \bibnamefont {Hudson}}, \bibinfo {author} {\bibfnamefont {S.~T.}\
  \bibnamefont {Flammia}}, \bibinfo {author} {\bibfnamefont {K.~M.}\
  \bibnamefont {Itoh}}, \bibinfo {author} {\bibfnamefont {A.}~\bibnamefont
  {Morello}}, \bibinfo {author} {\bibfnamefont {S.~D.}\ \bibnamefont
  {Bartlett}}, \ and\ \bibinfo {author} {\bibfnamefont {A.~S.}\ \bibnamefont
  {Dzurak}},\ }\href {\doibase 10.1038/s41928-019-0234-1} {\bibfield  {journal}
  {\bibinfo  {journal} {Nature Electronics}\ }\textbf {\bibinfo {volume} {2}},\
  \bibinfo {pages} {151} (\bibinfo {year} {2019})}\BibitemShut {NoStop}%
\bibitem [{\citenamefont {Chan}\ \emph {et~al.}(2018)\citenamefont {Chan},
  \citenamefont {Huang}, \citenamefont {Yang}, \citenamefont {Hwang},
  \citenamefont {Hensen}, \citenamefont {Tanttu}, \citenamefont {Hudson},
  \citenamefont {Itoh}, \citenamefont {Laucht}, \citenamefont {Morello},\ and\
  \citenamefont {Dzurak}}]{chan2018assessment}%
  \BibitemOpen
  \bibfield  {author} {\bibinfo {author} {\bibfnamefont {K.~W.}\ \bibnamefont
  {Chan}}, \bibinfo {author} {\bibfnamefont {W.}~\bibnamefont {Huang}},
  \bibinfo {author} {\bibfnamefont {C.~H.}\ \bibnamefont {Yang}}, \bibinfo
  {author} {\bibfnamefont {J.~C.~C.}\ \bibnamefont {Hwang}}, \bibinfo {author}
  {\bibfnamefont {B.}~\bibnamefont {Hensen}}, \bibinfo {author} {\bibfnamefont
  {T.}~\bibnamefont {Tanttu}}, \bibinfo {author} {\bibfnamefont {F.~E.}\
  \bibnamefont {Hudson}}, \bibinfo {author} {\bibfnamefont {K.~M.}\
  \bibnamefont {Itoh}}, \bibinfo {author} {\bibfnamefont {A.}~\bibnamefont
  {Laucht}}, \bibinfo {author} {\bibfnamefont {A.}~\bibnamefont {Morello}}, \
  and\ \bibinfo {author} {\bibfnamefont {A.~S.}\ \bibnamefont {Dzurak}},\
  }\href {\doibase 10.1103/PhysRevApplied.10.044017} {\bibfield  {journal}
  {\bibinfo  {journal} {Physical Review Applied}\ }\textbf {\bibinfo {volume}
  {10}},\ \bibinfo {pages} {044017} (\bibinfo {year} {2018})}\BibitemShut
  {NoStop}%
\bibitem [{\citenamefont {Elzerman}\ \emph {et~al.}(2004)\citenamefont
  {Elzerman}, \citenamefont {Hanson}, \citenamefont {{Willems van Beveren}},
  \citenamefont {Witkamp}, \citenamefont {Vandersypen},\ and\ \citenamefont
  {Kouwenhoven}}]{elzerman2004singleshot}%
  \BibitemOpen
  \bibfield  {author} {\bibinfo {author} {\bibfnamefont {J.~M.}\ \bibnamefont
  {Elzerman}}, \bibinfo {author} {\bibfnamefont {R.}~\bibnamefont {Hanson}},
  \bibinfo {author} {\bibfnamefont {L.~H.}\ \bibnamefont {{Willems van
  Beveren}}}, \bibinfo {author} {\bibfnamefont {B.}~\bibnamefont {Witkamp}},
  \bibinfo {author} {\bibfnamefont {L.~M.~K.}\ \bibnamefont {Vandersypen}}, \
  and\ \bibinfo {author} {\bibfnamefont {L.~P.}\ \bibnamefont {Kouwenhoven}},\
  }\href {\doibase 10.1038/nature02693} {\bibfield  {journal} {\bibinfo
  {journal} {Nature}\ }\textbf {\bibinfo {volume} {430}},\ \bibinfo {pages}
  {431} (\bibinfo {year} {2004})}\BibitemShut {NoStop}%
\bibitem [{\citenamefont {Yan}\ \emph {et~al.}(2013)\citenamefont {Yan},
  \citenamefont {Gustavsson}, \citenamefont {Bylander}, \citenamefont {Jin},
  \citenamefont {Yoshihara}, \citenamefont {Cory}, \citenamefont {Nakamura},
  \citenamefont {Orlando},\ and\ \citenamefont
  {Oliver}}]{yan2013rotatingframe}%
  \BibitemOpen
  \bibfield  {author} {\bibinfo {author} {\bibfnamefont {F.}~\bibnamefont
  {Yan}}, \bibinfo {author} {\bibfnamefont {S.}~\bibnamefont {Gustavsson}},
  \bibinfo {author} {\bibfnamefont {J.}~\bibnamefont {Bylander}}, \bibinfo
  {author} {\bibfnamefont {X.}~\bibnamefont {Jin}}, \bibinfo {author}
  {\bibfnamefont {F.}~\bibnamefont {Yoshihara}}, \bibinfo {author}
  {\bibfnamefont {D.~G.}\ \bibnamefont {Cory}}, \bibinfo {author}
  {\bibfnamefont {Y.}~\bibnamefont {Nakamura}}, \bibinfo {author}
  {\bibfnamefont {T.~P.}\ \bibnamefont {Orlando}}, \ and\ \bibinfo {author}
  {\bibfnamefont {W.~D.}\ \bibnamefont {Oliver}},\ }\href {\doibase
  10.1038/ncomms3337} {\bibfield  {journal} {\bibinfo  {journal} {Nature
  Communications}\ }\textbf {\bibinfo {volume} {4}},\ \bibinfo {pages} {2337}
  (\bibinfo {year} {2013})}\BibitemShut {NoStop}%
\bibitem [{\citenamefont {Zeng}\ \emph {et~al.}(2019)\citenamefont {Zeng},
  \citenamefont {Yang}, \citenamefont {Dzurak},\ and\ \citenamefont
  {Barnes}}]{zeng2019geometric}%
  \BibitemOpen
  \bibfield  {author} {\bibinfo {author} {\bibfnamefont {J.}~\bibnamefont
  {Zeng}}, \bibinfo {author} {\bibfnamefont {C.~H.}\ \bibnamefont {Yang}},
  \bibinfo {author} {\bibfnamefont {A.~S.}\ \bibnamefont {Dzurak}}, \ and\
  \bibinfo {author} {\bibfnamefont {E.}~\bibnamefont {Barnes}},\ }\href
  {\doibase 10.1103/PhysRevA.99.052321} {\bibfield  {journal} {\bibinfo
  {journal} {Physical Review A}\ }\textbf {\bibinfo {volume} {99}},\ \bibinfo
  {pages} {052321} (\bibinfo {year} {2019})}\BibitemShut {NoStop}%
\bibitem [{\citenamefont {Nielsen}\ and\ \citenamefont
  {Chuang}(2010)}]{nielsen2010quantum}%
  \BibitemOpen
  \bibfield  {author} {\bibinfo {author} {\bibfnamefont {M.~A.}\ \bibnamefont
  {Nielsen}}\ and\ \bibinfo {author} {\bibfnamefont {I.~L.}\ \bibnamefont
  {Chuang}},\ }\href@noop {} {\emph {\bibinfo {title} {Quantum Computation and
  Quantum Information: 10th Anniversary Edition}}}\ (\bibinfo  {publisher}
  {Cambridge University Press},\ \bibinfo {year} {2010})\BibitemShut {NoStop}%
\bibitem [{\citenamefont {Laucht}\ \emph {et~al.}(2016)\citenamefont {Laucht},
  \citenamefont {Simmons}, \citenamefont {Kalra}, \citenamefont {Tosi},
  \citenamefont {Dehollain}, \citenamefont {Muhonen}, \citenamefont {Freer},
  \citenamefont {Hudson}, \citenamefont {Itoh}, \citenamefont {Jamieson},
  \citenamefont {McCallum}, \citenamefont {Dzurak},\ and\ \citenamefont
  {Morello}}]{laucht2016breaking}%
  \BibitemOpen
  \bibfield  {author} {\bibinfo {author} {\bibfnamefont {A.}~\bibnamefont
  {Laucht}}, \bibinfo {author} {\bibfnamefont {S.}~\bibnamefont {Simmons}},
  \bibinfo {author} {\bibfnamefont {R.}~\bibnamefont {Kalra}}, \bibinfo
  {author} {\bibfnamefont {G.}~\bibnamefont {Tosi}}, \bibinfo {author}
  {\bibfnamefont {J.~P.}\ \bibnamefont {Dehollain}}, \bibinfo {author}
  {\bibfnamefont {J.~T.}\ \bibnamefont {Muhonen}}, \bibinfo {author}
  {\bibfnamefont {S.}~\bibnamefont {Freer}}, \bibinfo {author} {\bibfnamefont
  {F.~E.}\ \bibnamefont {Hudson}}, \bibinfo {author} {\bibfnamefont {K.~M.}\
  \bibnamefont {Itoh}}, \bibinfo {author} {\bibfnamefont {D.~N.}\ \bibnamefont
  {Jamieson}}, \bibinfo {author} {\bibfnamefont {J.~C.}\ \bibnamefont
  {McCallum}}, \bibinfo {author} {\bibfnamefont {A.~S.}\ \bibnamefont
  {Dzurak}}, \ and\ \bibinfo {author} {\bibfnamefont {A.}~\bibnamefont
  {Morello}},\ }\href {\doibase 10.1103/PhysRevB.94.161302} {\bibfield
  {journal} {\bibinfo  {journal} {Physical Review B}\ }\textbf {\bibinfo
  {volume} {94}},\ \bibinfo {pages} {161302} (\bibinfo {year}
  {2016})}\BibitemShut {NoStop}%
\bibitem [{\citenamefont {Feng}\ \emph {et~al.}(2016)\citenamefont {Feng},
  \citenamefont {Wallman}, \citenamefont {Buonacorsi}, \citenamefont {Cho},
  \citenamefont {Park}, \citenamefont {Xin}, \citenamefont {Lu}, \citenamefont
  {Baugh},\ and\ \citenamefont {Laflamme}}]{feng2016estimating}%
  \BibitemOpen
  \bibfield  {author} {\bibinfo {author} {\bibfnamefont {G.}~\bibnamefont
  {Feng}}, \bibinfo {author} {\bibfnamefont {J.~J.}\ \bibnamefont {Wallman}},
  \bibinfo {author} {\bibfnamefont {B.}~\bibnamefont {Buonacorsi}}, \bibinfo
  {author} {\bibfnamefont {F.~H.}\ \bibnamefont {Cho}}, \bibinfo {author}
  {\bibfnamefont {D.~K.}\ \bibnamefont {Park}}, \bibinfo {author}
  {\bibfnamefont {T.}~\bibnamefont {Xin}}, \bibinfo {author} {\bibfnamefont
  {D.}~\bibnamefont {Lu}}, \bibinfo {author} {\bibfnamefont {J.}~\bibnamefont
  {Baugh}}, \ and\ \bibinfo {author} {\bibfnamefont {R.}~\bibnamefont
  {Laflamme}},\ }\href {\doibase 10.1103/PhysRevLett.117.260501} {\bibfield
  {journal} {\bibinfo  {journal} {Physical Review Letters}\ }\textbf {\bibinfo
  {volume} {117}},\ \bibinfo {pages} {260501} (\bibinfo {year}
  {2016})}\BibitemShut {NoStop}%
\bibitem [{\citenamefont {Huang}\ \emph {et~al.}(2019)\citenamefont {Huang},
  \citenamefont {Yang}, \citenamefont {Chan}, \citenamefont {Tanttu},
  \citenamefont {Hensen}, \citenamefont {Leon}, \citenamefont {Fogarty},
  \citenamefont {Hwang}, \citenamefont {Hudson}, \citenamefont {Itoh},
  \citenamefont {Morello}, \citenamefont {Laucht},\ and\ \citenamefont
  {Dzurak}}]{huang2019fidelity}%
  \BibitemOpen
  \bibfield  {author} {\bibinfo {author} {\bibfnamefont {W.}~\bibnamefont
  {Huang}}, \bibinfo {author} {\bibfnamefont {C.~H.}\ \bibnamefont {Yang}},
  \bibinfo {author} {\bibfnamefont {K.~W.}\ \bibnamefont {Chan}}, \bibinfo
  {author} {\bibfnamefont {T.}~\bibnamefont {Tanttu}}, \bibinfo {author}
  {\bibfnamefont {B.}~\bibnamefont {Hensen}}, \bibinfo {author} {\bibfnamefont
  {R.~C.~C.}\ \bibnamefont {Leon}}, \bibinfo {author} {\bibfnamefont {M.~A.}\
  \bibnamefont {Fogarty}}, \bibinfo {author} {\bibfnamefont {J.~C.~C.}\
  \bibnamefont {Hwang}}, \bibinfo {author} {\bibfnamefont {F.~E.}\ \bibnamefont
  {Hudson}}, \bibinfo {author} {\bibfnamefont {K.~M.}\ \bibnamefont {Itoh}},
  \bibinfo {author} {\bibfnamefont {A.}~\bibnamefont {Morello}}, \bibinfo
  {author} {\bibfnamefont {A.}~\bibnamefont {Laucht}}, \ and\ \bibinfo {author}
  {\bibfnamefont {A.~S.}\ \bibnamefont {Dzurak}},\ }\href {\doibase
  10.1038/s41586-019-1197-0} {\bibfield  {journal} {\bibinfo  {journal}
  {Nature}\ }\textbf {\bibinfo {volume} {569}},\ \bibinfo {pages} {532}
  (\bibinfo {year} {2019})}\BibitemShut {NoStop}%
\end{thebibliography}%

\end{document}